\documentclass[runningheads]{llncs}
\usepackage[utf8]{inputenc}
\usepackage{adjustbox}

\usepackage{xcolor}
\usepackage{graphicx}
\usepackage{booktabs}
\usepackage{amsmath}

\usepackage{booktabs}
\usepackage{pgfplotstable}
\usepackage{rotating}
\usepackage{comment}

\usepackage{pgfplots}
\usepackage{pgfplotstable}

\usepackage{listings}

\usepackage{hyperref}
\usepackage{subcaption}
\usepackage{todonotes}

\let\llncssubparagraph\subparagraph
\let\subparagraph\paragraph
\usepackage[normalem]{ulem}
\let\subparagraph\llncssubparagraph

\begin{document}
\title{
Analyzing, Exploring, and Visualizing Complex Networks via Hypergraphs using SimpleHypergraphs.jl\thanks{The research is financed by NAWA --- The Polish National Agency for Academic Exchange.}}

\author{
Alessia Antelmi \inst{1} \and
Gennaro Cordasco \inst{2} \and
Bogumił Kamiński\inst{3} \and
Paweł Prałat \inst{4} \and
Vittorio Scarano \inst{2} \and
Carmine Spagnuolo \inst{2}\and
Przemyslaw Szufel\inst{3}}
\authorrunning{A. Antelmi et al.}
\titlerunning{SimpleHypergraphs.jl}

\institute{Dipartimento di Informatica, Università degli Studi di Salerno, Italy
\email{aantelmi@unisa.it,vitsca@unisa.it,cspagnuolo@unisa.it} \and
Dipartimento di Psicologia, Università degli Studi della Campania “Luigi Vanvitelli”, Italy
\email{gennaro.cordasco@unicampania.it} \and
SGH Warsaw School of Economics, Poland
\email{bkamins@sgh.waw.pl, pszufe@sgh.waw.pl} \and
Department of Mathematics, Ryerson University, Toronto, ON, Canada
\email{pralat@ryerson.ca} }

\maketitle

\begin{abstract}
\vspace{-20pt}
Real-world complex networks are usually being modeled as graphs. The concept of graphs assumes that the relations within the network are binary (for instance, between pairs of nodes); however, this is not always true for many real-life scenarios, such as peer-to-peer communication schemes, paper co-authorship, or social network interactions. For such scenarios, it is often the case that the underlying network is better and more naturally modeled by hypergraphs. A hypergraph is a generalization of a graph in which a single (hyper)edge can connect any number of vertices. Hypergraphs allow modelers to have a complete representation of multi-relational (many-to-many) networks; hence, they are extremely suitable for analyzing and discovering more subtle dependencies in such data structures. 

Working with hypergraphs requires new software libraries that make it possible to perform operations on them, from basic algorithms (such as searching or traversing the network) to computing significant hypergraph measures, to including more challenging algorithms (such as community detection). In this paper, we present a new software library, \texttt{SimpleHypergraphs.jl}, written in the Julia language and designed for high-performance computing on hypergraphs and propose two new algorithms for analyzing their properties: $s$-betweenness and modified label propagation.
We also present various approaches for hypergraph visualization integrated into our tool. In order to demonstrate how to exploit the library in practice, we discuss two case studies based on the 2019 Yelp Challenge dataset and the collaboration network built upon the Game of Thrones TV series. The results are promising and they confirm the ability of hypergraphs to provide more insight than standard graph-based approaches.

% along with a second example analyzing social network within the Game of Thrones series.
\end{abstract}

\section{Introduction}
Research on the analysis of networks has a long tradition, and have provided mathematics and computer scientists tools enabling the exploration, the study, and the comprehension of complex phenomena~\cite{Scharnhorst2006}. Since its birth in the Eighteenth century at the hands of the Swiss mathematician Leonhard Euler, \textit{graph theory} --- the branch of discrete mathematics dealing with the study of \textit{networks} --- has contributed to the resolution of many real-world problems~\cite{graphTheory,Fortunato2010}. In particular, over the last twenty years, the interests of research have focused on \textit{complex networks}, namely networks whose structure is irregular, complex and dynamically evolving in time~\cite{boccaletti2006,Interdonato2019,Strogatz2001}. Complex networks naturally model many real-world scenarios, such as social interactions~\cite{Chen2019,Nguyen2020}, biological~\cite{Hasenjager2020,Hossain2020,Romero2020} and economical~\cite{Gobel2020,Verba2020} systems, Internet~\cite{Gan2014}, and the World Wide Web~\cite{Scharnhorst2006}, just to name a few examples. Traditionally, these networks are described using graphs, where nodes represent elements of the network, and edges represent relationships between some pairs of elements. However, in many practical applications, relationships between the elements of a network may not be dyadic but may involve more than two nodes. Examples of such scenarios include membership in groups on social platforms, co-authorships of scientific publications, or several parties participating in a crypto-currency transaction~\cite{Bretto:2013:HTI:2500991}.
In such cases, nodes may be linked together based either on explicit information (e.g., inclusion in groups), or implicit information (e.g., whether online social network users share the same hashtag in a media post or review the same restaurant). Obviously, the resulting complexity of these networks is tremendous, as the relationships between vertices can involve an arbitrary number of elements. A challenging task arising in this context is providing scientists a tool to effortlessly model such scenarios. Here, \textit{hypergraphs} come into play. A hypergraph is a generalization of a graph where the vertices are related not only by pair-wise connections (edges), but they can include an arbitrary number of nodes (hyperedges). In other words, hypergraphs can naturally model all the above scenarios.

The powerful expressiveness of hypergraphs has, however, few drawbacks: dealing with the complexity of such data structures and the lack of appropriate tools and algorithms for their study. For this reason, hypergraphs have been little used in literature in favor of their graph-counterpart. A traditional approach in network science to handle such scenarios is using the \emph{two}-section graph representation of a hypergraph, which vertices are the vertices of the hypergraph and where two distinct vertices form an edge if and only if they are in the same hyperedge~\cite{Peng2017,Vargas2018}. In other words, a complete graph (or a clique) of order $k$ replaces each hyperedge of cardinality $k$.
Another way to deal with a hypergraph is analyzing its \textit{line}-graph, defined as the graph where the node-set is the set of the hyperedges and two nodes are connected by a link when the corresponding hyperedges share at least a node~\cite{Liu2013}. A third approach consists in using a bipartite graph, where the vertices and hyperedges of a hypergraph represent the two disjoint vertex sets. Recommender systems heavily manipulate such representation~\cite{Beel2016,Silveira2017,MUSTO2017}. 
However, all these techniques share a weakness: as they do not exploit hypergraphs, their implementation requires a different and less natural data structure to handle the same set of information. Additionally, both the two-section and line-graph transformations lose information encoded in a hypergraph that cannot be transferred to the corresponding graph. For a more clarifying example, we can consider the network built upon e-mail exchanges between some users. In this context, the object \textit{e-mail} can be modeled as a relation involving a group of users. Thus, in this case, nodes of the network represent the persons, while the edges of the network incorporate a sub-set of them -- i.e., all e-mail receivers. It is worth noting that if we represent this scenario with a graph, we lose the information about which users are receivers of the same e-mails. This approach, combined with grouping messages having the same title, can be used for anomaly and spam detection in electronic communication~\cite{silva2008hypergraph}.

To illuminate this uncharted area, we delved into the study of hypergraphs, discussing how and to what extent this mathematical structure can model, analyze, and visualize complex networks characterized by many-to-many relations. 
In this paper, we propose \texttt{SimpleHypergraphs.jl}, a complete software tool written in the Julia language. Here, our aim is two-fold: i) improving the usability and efficiency of software libraries for hypergraphs manipulation by exploiting the efficiency provided by Julia and ii) developing a holistic set of functionalities ensuring the broad applicability of our library. 
The contributions of our work can be summarized as follows:

\begin{itemize}
\item We propose a software library for the analysis, exploration, and visualization of hypergraphs, exploiting the Julia language to ensure both efficiency and expressiveness. Julia is a new programming language developed at MIT~\cite{bezanson2017julia}, with a syntax similar to popular and easy-to-use scientific computing languages such as Python or R. This means that experience in those languages can be directly applied in Julia by computational scientists~\cite{edelman2019julia,regier2019cataloging}. Although it keeps a math-oriented syntax, Julia compiles the code to a binary form. As a result, the observed performance of Julia programs is very similar to C++, but with around $4$ times fewer lines of code.

\texttt{SimpleHypergraphs.jl} is available on a GitHub public repository\footnote{\url{https://github.com/pszufe/SimpleHypergraphs.jl}.}, where it is possible to find the library documentation\footnote{\url{https://pszufe.github.io/SimpleHypergraphs.jl/latest/reference/}.}, and several tutorials in the form of Jupyter Notebooks\footnote{\url{https://github.com/pszufe/SimpleHypergraphs.jl/tree/master/examples}.}. In this article, we describe the library functionalities available in the current version $0.1.7$ of \texttt{SimpleHypergraphs.jl}. The library provides a set of analytical functionalities (modularity, connected components, random-walk), as well as a serialization mechanism to store hypergraph metadata. It also includes a visualization component that allows users to explore the network through two different hypergraph visualizations;

\item We discuss two use cases where we use hypergraphs to analyze complex networks, and we compare their performance with the corresponding two-section graph.  The first case study deals with business reviews from the platform \textit{Yelp.com}, while the second application investigates the relationships between characters of the \textit{Game-of-Thrones} TV Series. In order to perform these analyses we propose two new algorithms for analyzing the properties of hypergraphs: $s$-betweenness and modified label propagation.
\end{itemize}

The paper is structured as follows. In Section~\ref{sec:review}, we start by motivating the introduction of a novel software library to analyze and explore hypergraphs, and we provide a review of existing available (currently maintained) software tools.
Section~\ref{sec:library} defines the notation used in this work and introduces our Julia-based library for hypergraphs. Next, in Section~\ref{sec:usecases}, we present two use cases with the aim to show concrete applications and discuss the $s$-betweenness and modified label propagation algorithms we propose. Finally, in Section~\ref{sec:conclusion}, we discuss some conclusions and future directions.

\section{Motivation}\label{sec:review}
Hypergraphs are a natural generalization of graphs, where a single (hyper)edge can connect more than two vertices. In several real-world applications, such representation is not only more general but also more natural than a standard graph representation, where a binary representation of relationships is sometimes not sufficient to correctly capture subtle interactions. Typical applications of hypergraphs include modeling paper co-authorship networks (different authors contribute to the same paper~\cite{han2009understanding}), online reviews (the same good purchased by several users~\cite{yuan2018semantic}), social network activities (the same post commented by multiple users, links in social networks~\cite{li2013link}), disease contingency plans (groups of people locating in the same place~\cite{bodo2016sis}), bio-engineering systems (modeling cellular networks~\cite{klamt2009hypergraphs}).
Even though hypergraphs are natural representations of many real-world systems, there currently are very few software frameworks suitable for modeling and mining these structures. In this section, we give a brief state-of-art overview of several software libraries, focusing on the availability of their code and their capability to model and analyze hypergraphs.

\begin{itemize}

\item \textbf{Chapel HyperGraph Library (CHGL)}~\cite{chapel} has been developed by the Pacific Northwest National Laboratory since 2018, and released under the MIT license. CHGL is a library for the emerging parallel language Chapel. It provides the \texttt{AdjListHyperGraph} module that allows storing hypergraphs on shared and distributed memory. The library is not well documented and does not support any easy mechanism for the two-section and bipartite analyses. Nonetheless, it is worth mentioning for its compatibility for parallel and distributed computing.

\item \textbf{HyperX}~\cite{hyperx} is a scalable framework for processing hypergraphs and learning algorithms built on top of Apache Spark. This library supports the same design model of GraphX, the Apache Spark API for graphs, and graph-parallel computation written in the Scala language. An interesting feature of this library is that it provides native support for the hypergraph elaboration. Directly processing hypergraph data, HyperX obtains significant speedup with respect using the bipartite or the two-section representation of a hypergraph and then exploiting the GraphX APIs.
%The standard approach uses the bipartite or the 2-section representation of hypergraphs and exploits the GraphX library, while HyperX directly processes hypergraph data obtaining significant speedup compared to the standard approach.

\item \textbf{Pygraph}~\cite{pygraph} is a pure Python library for graph manipulation, released under the MIT license. It supports a hypergraph representation by exposing the class \texttt{hypergraph}, even though it does not provide any specific optimization or functionality for hypergraphs.
%It implements almost all basic functionalities on graphs, but also supports hypergraphs by exposing the class \texttt{hypergraph}. This library does not provide any specific optimization or functionalities for hypergraphs.

\item \textbf{Multihypergraph}~\cite{multihypergraph} is a Python package for graphs, released under the GPL license. The library emphasizes the mathematical understanding of graphs rather than the algorithmic efficiency, and it provides support for hyper-edges, multi-edges, and looped-edges. This library implements only the graph memory model definition and isomorphism functionalities, without defining any other functionality and algorithm for graphs and hypergraphs. %In this library are not included any kind of functionalities for hypergraph and does not provide algorithms for hypergraph or graph, but only graph model memory definition and isomorphism functionalities.

\item \textbf{HyperNetX}~\cite{HyperNetX} is a Python preliminary library released in 2018 under the Battelle Memorial Institute license. This library generalizes traditional graph metrics (such as vertex and edge degrees, diameter, distances) to hypergraphs, and it provides proper documentation and tutorials.  HyperNetX supports the bipartite representation of a hypergraph, along with the possibility to load hypergraphs from their bipartite view. Furthermore, it exports some visualization functionalities for hypergraphs based on Euler-diagrams.
%\footnote{https://github.com/pnnl/HyperNetX/blob/master/LICENSE.rst}

\item \textbf{Halp}~\cite{halp} is a Python software package providing both a directed and an undirected hypergraph implementation, as well as several major and classical algorithms. The library is developed by Murali's Research Group at Virginia Tech, and it is released under the GPL license. The library provides several statistics on hypergraphs and model transformations in graphs, supported by the \texttt{NetworkX} Python library. Several algorithms for hypergraphs, such as $k$-shortest-hyperpaths and random walk, are also implemented.

\item \textbf{HyperGraphLib}~\cite{HyperGraphLib} is a C++ implementation of hypergraphs that exploits the Boost Library, also defining the library license. This library provides basic functionalities for hypergraphs and implements some simple metrics. Moreover, it provides isomorphism functionalities and path-finding algorithms. However, it does not implement any hypergraph representation into a graph (such as a bipartite or a 2-section graph) nor software integration with other graph libraries.

\item \textbf{Iper}~\cite{iper} is a JavaScript library for hypergraphs, released under the MIT license. The library defines a hypergraph and allows the user to define meta-information for vertices. However, it does not include any hypergraph transformation and integration with other graph libraries for classical statistics and algorithms.

\item \textbf{NetworkR}~\cite{networkR} is an R package with a set of functions for analyzing social and economic networks, including hypergraphs. It incorporates analyses such as degree distribution, and density of the network, as well as microscopic level analysis such as power, influence, and centrality of individual nodes. The library does not provide support for meta-information on vertices and hyperedges and provides only hypergraphs projection into graphs.

\item \textbf{Gspbox}~\cite{gspbox} is an easy to use Matlab toolbox that performs a wide variety of operations on a graph. It is based on spectral graph theory, and many of the implemented features can scale to very large graphs. Gspbox supports hypergraphs modeling, including weighted hyperedges, and vertices with coordinates in the space. The hypergraph manipulation is obtained by representing the model as a graph. For this reason, although all graph functionalities are available, the library does not provide any specific solutions or optimization for hypergraphs.

\end{itemize}
Overall, all the considered libraries are a compromise between efficiency, which characterizes low-level languages such as C/C++, and the easy-of-use and expressiveness, distinguishing interpreted and scripting languages such as Python and R.  

%Formally,  a graph (undirected) is an ordered pair $G = (V, E)$ having $n = |V|$ vertices and $m = |E|$ edges, which are $2$-element subsets of $V$. For each $v \in V$, we denote by $N(v) = \{u: u \in V, \{u, v\} \in E\}$ the neighbourhood of $v$ and by $\text{deg}(v) = | N(v)|$ the degree of $v$.

%While in a graph each edge joins exacty two vertices in an hypergraph each edge can join any number of vertices. Formally, a hypergraph $H$  is a pair $H=(V,E)$, having $n = |V|$ vertices, and  $k=|E|$ hyperedges, which are  non-empty subsets of $V$. Therefore, $E$ is a subset of $\mathcal{P}(V)\setminus \{\emptyset \}$, where ${\mathcal P}(X)$ denotes the power set of a set $X$.

\section {SimpleHypergraphs.jl}\label{sec:library}
In this Section, we present the \texttt{SimpleHypergraphs.jl} library, which provides flexible functionalities for the analysis and modeling of hypergraphs. Being implemented in the Julia programming language, and released under the terms of the Open Source MIT License, it is currently part of the official Julia package repository. This section is organized as follows. We introduce the adopted formal notation and definitions, and we then move to describe the library design and memory model. Finally, we discuss its functionalities about hypergraph manipulation, analysis, and visualization. 

\subsection{Definitions and notation} \label{sec:notation}
Formally, a hypergraph~\cite{Bretto:2013:HTI:2500991} is an ordered pair $H=(V,E)$ where $V$ is the set of nodes (often also called vertices) and $E$ is the set of edges. Each edge is a non-empty subset of vertices; i.e., $E \subseteq 2^V \setminus \{\emptyset\}$, where $2^V$ is the power set of $V$. We use $n=|V|$ and $m=|E|$ to indicate the size of the vertex set and the edge set, respectively. A graph can be seen as a hypergraph where each hyperedge is a two element subset of $V$. In other words, a graph $G=(V,E)$ is a hypergraph, if $E \subseteq \binom{V}{2} \subseteq 2^V \setminus \{\emptyset\}$. 

%Indeed, hypergraphs are generalization of graphs in which each edge is a two element subset of V

\subsection{Library design}
\texttt{SimpleHypergraphs.jl} provides APIs representing a hypergraph $H=(V,E)$ as an $n \times k$ matrix, where $n$ is the number of vertices and $k$ is the number of hyperedges. In other words, each row of the matrix is associated with a vertex and indicates the hyperedges the vertex belongs to.
In the APIs, vertices and hyperedges are uniquely identified by progressive integer ids, corresponding to rows ($1,\ldots, n$) and columns ($1,\ldots, k$), respectively.
Each position ($i,j$) of the matrix denotes the weight of the vertex $i$ within the hyperedge $j$.
The library also provides several constructors for defining meta-information type and enables to attach meta-data values of arbitrary type to both vertices and hyperedges.
To ensure flexible co-operability, \texttt{SimpleHypergraphs.jl} provides two-fold integration both with Julia standard matrix and \texttt{LightGraphs.jl} APIs.
%To ensure flexible co-operability, \texttt{SimpleHypergraphs.jl} provides two-fold integration with Julia standard matrix APIs and the APIs of \texttt{LightGraphs.jl}.

\begin{itemize}
    \item [i.] \textit{Julia's matrix APIs}.
    We achieved the Julia \texttt{Array} APIs integration by making the \texttt{Hypergraph} struct a subclass of \texttt{AbstractMatrix}, and providing a set of integration methods for manipulating matrices (i.e. querying the matrix size, fetching/updating elements). Internally, a hypergraph is stored as a sparse array. %; its data are held in a hashmap structure. 
    Hypergraph data are stored in a redundant format, using two separate hashmap structures for rows and columns to ensure good algorithmic performances. This design choice simultaneously provides high-grade performance across rows and columns. Furthermore, it avoids the circumstance where all data need to be rewritten when the adjacency matrix is updated (typical disadvantage of a compressed sparse row matrix). 
    As a subclass of \texttt{AbstractMatrix}, the hypergraph adjacency matrix can be manipulated just like any other matrix in Julia. As a result, from the user's point of view, a hypergraph $H=(V,E)$ can be seen as a $n\times k$ matrix representation, where $n$ is the number of vertices and $k$ is the number of hyperedges. 
    %In other words, the rows of the matrix are vertices, while columns are hyperedges. 
    Vertices and hyperedges are uniquely identified by progressive integers, corresponding to rows ($1,\ldots,n$) and columns ($1,\ldots,k$), respectively.
    Moreover, the library supports generic type metadata for both vertices and hyperedges.
    
    \item [ii.] \textit{LightGraphs.jl}.
    We obtained the integration with this Julia library to manipulate graphs by creating hypergraph ``view'' classes providing a representation of a hypergraph as either a bipartite or a two-section graph. Those representations actually do \emph{not} copy the data, but provide a view (\texttt{TwoSectionView} and \texttt{BipartiteView}) that allow to access the hypergraph data in a read-only mode. 
    As we developed a full set of integration methods for the \texttt{LightGraphs.jl} library, the user can directly analyze a hypergraph structure with all functionality provided by \texttt{LightGraphs.jl}.
\end{itemize}

\subsection{Memory model and  functionalities}
The latest release $0.1.7$ of \texttt{SimpleHypergraphs.jl} provides a range of new functionalities and methods to build and explore hypergraphs. The following sections introduce the hypergraph representation, several basics operations, and transformations, the serialization mechanisms (raw and JSON formats), a set of analytical algorithms (Section~\ref{ssec:analytical}), and two visualization strategies (Section~\ref{ssec:hg_viz}).

\subsubsection{Hypergraph constructors.}
The Julia hypergraph object is defined as:

\begin{center}
\begin{lstlisting}
Hypergraph{T, V, E, D} <: AbstractMatrix{Union{T, Nothing}}
\end{lstlisting}
\end{center}

%\vspace*{-0.2truecm}
where \texttt{T}, a subtype of \texttt{Real}, represents the type of the weights stored in the structure; \texttt{V} and \texttt{E} are the types of the meta-data values stored in the vertices and edges of the hypergraph, respectively; and \texttt{D}, a subtype of \texttt{AbstractDict\{Int,T\}}, is the type of the underlying  dictionary used for storing the weight values. Note that when calling the constructor, the parameters \texttt{\{T, V, E, D\}} can be omitted (starting from the rightmost). The default value for the dictionary type \texttt{D} is a standard Julia dictionary \texttt{Dict\{Int, T\}} (where \texttt{T} is the type of the weights). The default value for vertex and edge metadata types, \texttt{V} and \texttt{E}, is \texttt{Nothing} --- i.e., by default, no metadata is stored. A new empty hypergraph can be built specifying the number of vertices (rows) and hyperedges (columns). Optionally, a hypergraph can be either materialized starting from a given matrix or a \texttt{LightGraphs.jl} graph object.

\subsubsection{Querying and manipulating functions.}
\texttt{SimpleHypergraphs.jl} provides several accessing and manipulating functions:

\begin{itemize}
    \item \texttt{add\_vertex!} adds a vertex to a given hypergraph $H$. Optionally, the vertex can be added to existing hyperedges. Additionally, a value can be stored with the vertex using the \texttt{vertex\_meta} keyword parameter.
    \item \texttt{remove\_vertex!} removes a vertex from a given hypergraph $H$.
    \item \texttt{set\_vertex\_meta!} sets a new meta-value $new\_value$ for vertex $id$ in $H$.
    \item \texttt{get\_vertex\_meta} returns the meta-value stored at vertex $id$ in $H$.
    \item \texttt{get\_vertices} returns the vertices for a given hyperedge $he_{id}$ in $H$.
    \item \texttt{nhv} returns the number of vertices in  the hypergraph $H$.
\end{itemize}
We implemented analogous functionalities for the hyperedges.

\subsubsection{Hypergraph transformations.}
The library provides two hypergraph transformations into the corresponding graph representations.

\begin{figure*}[t!]
    \centering
    \begin{subfigure}[t]{0.5\textwidth}
        \includegraphics[width=\textwidth]{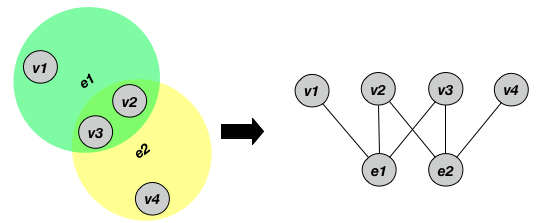}
        \caption{$G$ - Bipartite view of $H$. \label{fig:hbipartite}}
    \end{subfigure}%
    ~
    \begin{subfigure}[t]{0.5\textwidth}

        \includegraphics[width=\textwidth]{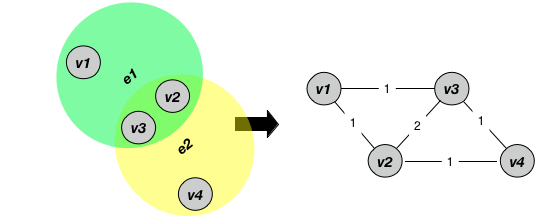}
        \caption{$G$ - Two-section view of $H$. \label{fig:h2section}}
    \end{subfigure}
    \caption{Hypergraph transformations into the graph counterpart.} %\vspace*{-0.2truecm}
\end{figure*}

\begin{enumerate}
    \item \texttt{BipartiteView}. It is the bipartite representation of a hypergraph $H$. As described in Bretto~\cite{Bretto:2013:HTI:2500991}, this representation is the incidence graph of the hypergraph $H=(V,E)$; that is, a bipartite graph $IG(H)$ with vertex set $S=V \cup E$, and where $v \in V$ and $e \in E$ are adjacent if and only if $v \in e$. Figure~\ref{fig:hbipartite} (on the left) illustrates a simple example of bipartite view.
    \item \texttt{TwoSectionView}. It is a two-section representation of a hypergraph $H$. As described in Bretto~\cite{Bretto:2013:HTI:2500991}, this representation of a hypergraph $H=(V,E)$, denoted with $[H]_2$, is a graph whose vertices are the vertices of $H$ and where two distinct vertices form an edge if and only if they are in the same hyperedge of $H$. As a result, each hyperedge from $H$ occurs as a complete graph in $G$. The weight of an edge corresponds to the number of hyperedges that contain both the endpoints of the edge.
    %\todo{PP: We need to say something about weights as this is what we use?}
    %\todo{PP: might be good to have an example with at least one hyperedge of size 3; that is, hypergraph that is not a graph. ;-)}
    Figure~\ref{fig:h2section} (on the right) details a trivial example of two-section view.
\end{enumerate}
Both \texttt{Views} are instances of \texttt{AbstractGraph}, the graph object defined by the \texttt{LightGraphs.jl} library. %\todo{PP: sometimes we use textit but sometimes texttt; we should be consistent.}
%library~\cite{LightGraphs}. 
When the view is materialized, according to the package specifics, the generated graph does not include any meta information.
%Both the transformations generates an \texttt{AbstractGraph} graph defined by the library \textit{LightGraphs.jl} \cite{LightGraphs}, according to the library' specific the generated graph does not include meta information.

\subsubsection{Hypergraph serialization.} 
The library currently offers two mechanisms to load and save a hypergraph from or to a stream. Given a hypergraph $H$, it may be stored using either a \textit{plain text} or \textit{JSON} formats.

\paragraph{Plain text format,} denoted by the \texttt{HGF\_Format} storage type. The first line consists of two integers $n$ and $k$, representing the number of vertices and the number of edges of $H$, respectively. The following $k$ rows (lines in a text file) describe the actual structure of $H$: each line represents a hyperedge and contains a list of all vertex-weight pairs within that hyperedge.

\paragraph{JSON format,} denoted by the \texttt{JSON\_Format} storage type. The internal hypergraph structure is represented with a dictionary that is serialized into a plain JSON object. Each dictionary key represents a hypergraph field, while each dictionary value stores the corresponding hypergraph field value. Additionally, the matrix view of $H$ is also stored. We used the Julia package \texttt{JSON3.jl} to handle the interaction between JSON and Julia types.

%\subsection{Performance benchmarks}

\subsection{Analytical functionalities} \label{ssec:analytical}
%MOVED TO THE START OF SECTION
%{\color{red} The library is constantly being upgraded and new functionalities are being added as needed. Below are some examples of build-in functionalities available at the time of writing this paper.} \todo{PP: OK to say this?}

\subsubsection{Hypergraph modularity.} 
Community detection is one of the most frequent tasks as it helps to find hidden interaction patterns in relational data. Newman and Girvan firstly proposed a hierarchical method to detect communities in complex systems introducing the concept of modularity~\cite{newman2004finding}. The modularity value is based on the comparison between the actual density of edges inside a community and the density one would expect to have if the vertices of the graph were attached at random. Higher modularity values signify denser connections between the nodes within clusters but more sparse connections between nodes in different clusters. In \texttt{SimpleHypergraphs.jl}, we implemented a generalization of the modularity notion, recently proposed for hypergraphs~\cite{kaminski2019modularity}. We further provided an algorithm to calculate the modularity of a given vertex partition.  %of a hypergraph 
This functionality is achieved via the \texttt{modularity} function which takes in input a hypergraph and its proposed partition given as a \verb|Vector{Set{Int}}| object.
% In \texttt{SimpleHypergraphs.jl}, we provide an implementation of an algorithm that allows to calculate the modularity of a given vertex partition of a hypergraph. Modularity was introduced by Newman and Girvan~\cite{newman2004finding}, and it is based on the comparison between the actual density of edges inside a community and the density one would expect to have if the vertices of the graph were attached at random. We implemented a generalization of the modularity notion that was recently proposed for hypergraphs in~\cite{kaminski2019modularity}. This functionality is provided via the \texttt{modularity} function that takes in input a hypergraph and its proposed partition given as a \verb|Vector{Set{Int}}| object.

\subsubsection{Connected components explorations.} 
The function \verb|get_connected|\\ \verb|_components| takes a hypergraph $H$ as input and returns a vector of vectors. Each vector represents a set of vertices that are a connected component in $H$. We define two vertices $a$ and $b$ of a hypergraph $H$ to be contained in a connected component if and only if $a$ and $b$ are connected by some sequences of hyperedges.

\subsubsection{Random walk.} 
Defining a random walk on a hypergraph is more complicated than defining the same notion on a graph, and there are a few natural ways it can be defined. One approach can be the following.  Suppose the walk starts from some vertex $i$. Then it can randomly select a hyperedge $i$ belongs to, and next choose a target vertex within that hyperedge, again at random. 
\texttt{SimpleHypergraphs.jl} provides the function \verb|random_walk| that takes a hypergraph and a starting vertex id as input and returns a destination vertex id in one step of the walk. This design choice guarantees full flexibility in defining random walks on hypergraphs. 
The function also accepts two optional keyword arguments, both functions: \texttt{heselect} and \texttt{vselect}. The first function specifies the rule by which hyperedge is selected for a given starting vertex. The second parameter selects the destination vertex from the selected hyperedge. By default, \texttt{heselect} chooses a hyperedge containing the source vertex uniformly at random. Similarly, \texttt{vselect} selects a vertex from a given hyperedge uniformly at random.

%\paragraph{\sout{Information Diffusion or Gossip?}}\todo{Gennaro and Przemyslaw. PSZ: Gennaro, since we have several pages already I propose to postpone this topic to the separate paper}

\subsection{Hypergraph visualization}\label{ssec:hg_viz}
\begin{figure*}[t!]
    \centering
    \begin{subfigure}[t]{0.5\textwidth}
        \includegraphics[width=\textwidth]{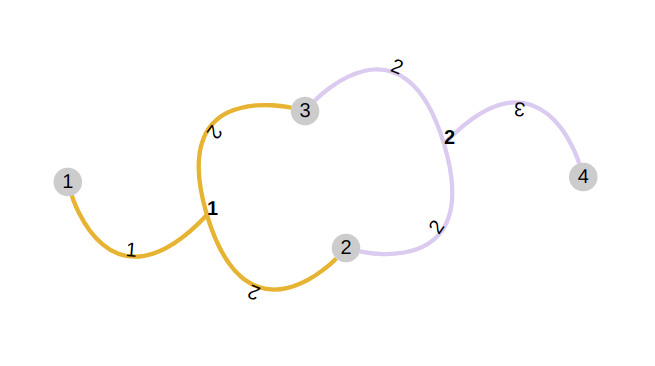}
        \caption{JavaScript visualization.\label{fig:js-vis}}
    \end{subfigure}%
    ~
    \begin{subfigure}[t]{0.5\textwidth}
        \includegraphics[width=\textwidth]{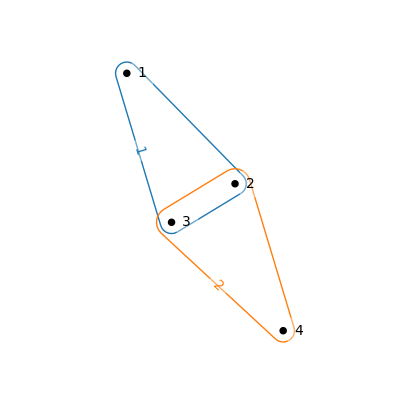}
        \caption{HyperNetX visualization. \label{fig:hypernetx}}
    \end{subfigure}
    \caption{\texttt{SimpleHypergraphs.jl} visualization methods. \label{fig:hg_vis}}
\end{figure*}

\texttt{SimpleHypergraphs.jl} currently offers the possibility to draw a hypergraph by exploiting two kinds of visualization, through the function \texttt{draw}. The available plotting methods are either based on an interactive JavaScript (JS) or a static Python-based solution. Figure~\ref{fig:hg_vis} illustrates the same hypergraph drawn using the two different strategies. A more detailed description follows. 

\paragraph{A JS-based visualization.} 
When dealing with complex objects that need to be visualized, it is of fundamental importance to have the possibility to easily catch the main information and, at the same time, to be able to retrieve more detail on demand~\cite{tufte83}. For this reason, we decided to integrate a dynamic and interactive visualization within \texttt{SimpleHypergraphs.jl}. This visualization is a wrapper around an external JS package that exploits \texttt{D3.js}, a JS library for manipulating documents based on data, which combines powerful visualization components and a data-driven approach to DOM manipulation. This architectural stack provides the user with a way to generate a dynamic visualization embeddable into a web-based environment, such as a Jupyter Notebook. This method represents each hyperedge $he$ of an hypergraph $H$ as a new fictitious vertex $fv$ to which each vertex $v \in he$ is connected (see Figure~\ref{fig:js-vis}). The appearance of vertices and hyperedges, whether displaying vertex weights and vertex and hyperedge metadata and labels are customizable.  

\paragraph{A Python-based visualization.}
This visualization is a wrapper around the drawing functionalities offered by the Python library \texttt{HyperNetX}~\cite{HyperNetX}, built upon the Python package \texttt{matplotlib}. \texttt{HyperNetX} renders a Euler diagram of the hypergraph where vertices are black dots and hyperedges are convex shapes containing the vertices belonging to the edge set (see Figure~\ref{fig:hypernetx}). As the authors note, it is not always possible to render the \textit{correct} Euler diagram for an arbitrary hypergraph. For this reason, this technique may lead to cases where a hyperedge incorrectly contains a vertex not belonging to its set. This library allows the user to manipulate the appearance of the resulting plot by letting the user defining the desired label, node, and edge options. \texttt{SimpleHypergraphs.jl} fully integrates the visualization potentiality of \texttt{HyperNetX}.

\section{Use cases}\label{sec:usecases}
In this Section, we discuss two use cases where we exploit hypergraphs to analyze complex data structures. The first case study deals with customers reviewing businesses on the social platform \textit{Yelp.com}, while the second application investigates the relationships between characters of the \textit{Game of Thrones} TV Series.

\subsection{Exploring and analyzing user reviews: \textit{Yelp.com}} \label{sec4}
In this first use case, we present a practical application of \texttt{SimpleHypergraphs.jl} applied to the analysis of business reviews from the online platform \textit{Yelp.com}~\cite{yelp}. A hypergraph is an accurate representation of such data, where vertices symbolize Yelp businesses and hyperedges costumers who reviewed multiple businesses. 
An attractive property that is worth investigating in this context is the community structure~\cite{newman2004finding}, i.e., the division of the network into groups of vertices that are similar among themselves but dissimilar from the rest of the network. The capability to detect the partitioning of a network into communities can give valuable insights into the organization and behavior of the system that the network models~\cite{ELMOUSSAOUI2019,Fortunato2010}.
In this particular case, the topology of the so-built hypergraph suggests clusters of businesses that users commonly review together. As hypergraph clustering~\cite{zhou2007learning} is an example of an unsupervised learning technique, our goal is to learn if such clusters are related to some natural characteristics of the underlying businesses. Such analysis allows us to understand which factors (ground-truth) influence the chance that a given user reviews any two businesses. To that end, we propose a methodology to measure and then to compare the results of hypergraph clustering against various possible ground-truth variables. In this context, the main challenge is to develop a measure comparable across different ground truths. 
Since the \textit{Yelp} dataset serves only as an example, the proposed approach can be used to identify ground-truths in other datasets representable as a hypergraph. 
As a side effect of this use case, we point out that the hypergraph based approach conveys more information about the ground-truth properties of a network than a standard graph analysis. In particular, we compare the results obtained for hypergraphs with the results achieved for the corresponding two-section graphs and show that hypergraph clusters provide uniformly more information than their graph counterpart. 
In more detail, we analyzed five different sub-hypergraphs, each one containing only reviews with the same number of stars, from 1 to 5. This approach shed some light on how to review linkages form; in particular, we studied how the mechanism behind those linkages differs across different review classes.

To summarize, we were interested in the following two research questions. i)~Does modeling the \textit{Yelp} dataset with hypergraphs give more qualitative information than looking at the corresponding two-section graph representation? ii)~Given the three hypergraphs consisting of positive, neutral, and negative reviews, are the these similar and to what extent? To answer these two questions, we set up the two experiments explained below.

\subsubsection{The \textit{Yelp.com} dataset.} 
\textit{Yelp} is an online platform where customers can share their experiences with local businesses by posting reviews, tips, photos, and videos. It allows businesses and customers to engage and transact~\cite{yelp}. Every year, the Yelp Inc.\ Company releases part of their data as an open dataset to grant the scientific community to conduct research and analysis on them. Some interesting articles that use the Yelp dataset for their analysis can be found in~\cite{Gulati2018,Ji2019,Li2019,Lu2018}. As a use case, we analyzed the 2019 Yelp Challenge dataset~\cite{yelp-dataset}, containing information about businesses, reviews, and users. Table~\ref{tab:tbl-yelp-fields} describes all the accessible dataset \textit{entities}. A more detailed description can be found on the official page~\cite{yelp-dataset-docs}.
%Fields contained in the dataset have been presented in Table \ref{tab:tbl-yelp-fields}.
%
\begin{table}[tb!]
    \centering
    \begin{tabular}{lcp{0.7\textwidth}}
        \toprule
            \textbf{Data} & \textbf{Instances} & \textbf{Description}\\
            \midrule
            Business & 192,609 & Business data including location, attributes, and categories.\\
            User & 1,637,138 & User data including the user's friend mapping and all the metadata associated with the user. \\
            Review & 6,685,900 & Full review text including the \texttt{user\_id} that wrote the review and the \texttt{business\_id} the review is written for. \\
            Picture & 200,000 & Photo data including caption and classification (one of ``food'', ``drink'', ``menu'', ``inside'' or ``outside'').\\
            Tip & 1,223,094 & Tips written by users on businesses. Tips are shorter than reviews and tend to convey quick suggestions.\\
            Check-in & 192,609 & Aggregated check-ins over time for each business. \\
        \bottomrule
        \vspace{5pt}
    \end{tabular}
      \caption{\textit{Yelp} entities contained in the dataset.  \label{tab:tbl-yelp-fields}}
\end{table}
Figure~\ref{fig:businesses-distribution} (on the left) presents business categories distribution, where a category is a label describing the typology of the business such as \textit{Bars} or \textit{Shopping} along with the number of reviews associated with each category.
It highlights the category distribution evaluated over all businesses. As clearly visible from the plot, the most common business typology is \textit{Restaurant}. For this reason, we focused our analysis on this business subgroup. Figure~\ref{fig:businesses-distribution} (on the right) shows the category distribution evaluated only within the \textit{Restaurant} macro-category. Both Figures show the top-20 most common categories. 
It is worth noting that as each Yelp restaurant may offer several types of cuisine (e.g., Indian, Chinese, Asian fusion), we labeled each business with a single \textit{food category}, assigning to them the most frequent tag in the database. Namely, if a restaurant \textit{R} had two genres \textit{A} and \textit{B}, but \textit{A} was overall more frequent in the dataset, we labeled \textit{R} with \textit{A}. 

%histogram about category distribution and #reviews per category
%\begin{figure}[ht!]
%    \centering
%  \includegraphics[width=1\linewidth]{businesses_reviews_distribution.pdf}
%  \caption{Businesses categories distribution and number of reviews associated with each category.}
  %improve label
%  \label{fig:businesses-distribution}
%\end{figure}

\pgfplotstableread[col sep=comma]{businesses.csv}\datatable
\pgfplotstableread[col sep=comma]{restaurants.csv}\datatabletwo
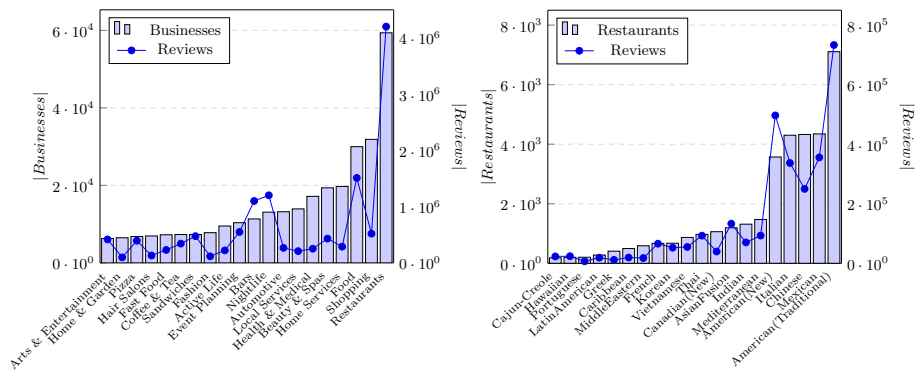
\begin{figure}[b!]
     % \hspace*{-1truecm}  
     \begin{subfigure}[t!]{.5\textwidth}
        \begin{tikzpicture}[scale=0.64] %[font=\small]
            \centering
            %restaurant distribution
            \begin{axis}[ %params
                    ybar,
                    ybar legend,
                    scale only axis,
                    width=\textwidth,%.5\linewidth,
                    ymajorgrids,
                    grid style={dashed,gray!30},
                    ylabel=$|Businesses|$,
                    axis y line*=left,
                    ymin=0,
                    ytick pos=left,
                    % yticklabel style={
                    %     /pgf/number format/fixed,
                    %     /pgf/number format/precision=5,
                    % },
                    y tick label style={/pgf/number format/sci},
                    scaled y ticks=false,
                    xtick=data,
                    /pgf/bar width=7pt,
                    enlarge x limits={0.027},
                    legend style={at={(0.5,-0.2)},anchor=north},
                    x tick label style={rotate=45,anchor=east,font=\scriptsize},
                    xticklabels from table={\datatable}{category},
                    xtick style={draw=none},
                    legend pos=north west
                ]
                \addplot[ybar, ybar legend, fill=blue!20]
                table[x expr=\coordindex,y=businesses] {\datatable};
                \label{businesses}
            \end{axis}
    
            %reviews count per category
            \begin{axis}[
                    width=\textwidth,%.5\linewidth,
                    scale only axis,
                    axis y line*=right,
                    ylabel near ticks,
                    ytick style={draw=none},
                    yticklabel pos=right,
                    % yticklabel style={
                    %     /pgf/number format/fixed,
                    %     /pgf/number format/precision=5,
                    % },
                    y tick label style={/pgf/number format/sci},
                    scaled y ticks=false,
                    ylabel=$|Reviews|$,
                    ylabel style={rotate=-180},
                    ymin=0,
                    ymax=4500000,
                    axis x line=none,
                    enlarge x limits={0.027},
                    legend pos=north west
                ]
                \addlegendimage{/pgfplots/refstyle=businesses}\addlegendentry{Businesses}
    
                \addplot +[draw=blue]
                table[x expr=\coordindex,y=reviews] {\datatable};
                \label{reviews}
    
                %\addlegendimage{/pgfplots/refstyle=reviews}
                \addlegendentry{Reviews}
            \end{axis}
        \end{tikzpicture}
    \end{subfigure}
    ~
    \begin{subfigure}[t!]{.5\textwidth}
     \begin{tikzpicture}[scale=0.64]
        %restaurant distribution
        \begin{axis}[ %params
                ybar,
                ybar legend,
                scale only axis,
                width=\textwidth,%.5\linewidth,
                ymajorgrids,
                grid style={dashed,gray!30},
                ylabel=$|Restaurants|$,
                axis y line*=left,
                ymin=0,
                ymax=8500,
                ytick pos=left,
                y tick label style={/pgf/number format/sci},
                xtick=data,
                /pgf/bar width=7pt,
                enlarge x limits={0.027},
                legend style={at={(0.5,-0.2)},anchor=north},
                x tick label style={rotate=43,anchor=east,font=\scriptsize},
                xticklabels from table={\datatabletwo}{category},
                xtick style={draw=none},
                legend pos=north west
            ]
            \addplot[ybar, ybar legend, fill=blue!20]
            table[x expr=\coordindex,y=businesses] {\datatabletwo};
            \label{restaurants}
        \end{axis}

        %reviews count per category
        \begin{axis}[
                width=\textwidth,%.5\linewidth,
                scale only axis,
                axis y line*=right,
                ylabel near ticks,
                ytick style={draw=none},
                yticklabel pos=right,
                % yticklabel style={
                %     /pgf/number format/fixed,
                %     /pgf/number format/precision=5,
                % },
                y tick label style={/pgf/number format/sci},
                scaled y ticks=false,
                ylabel=$|Reviews|$,
                ylabel style={rotate=-180},
                ymin=0,
                ymax=850000,
                axis x line=none,
                enlarge x limits={0.027},
                legend pos=north west
            ]
            \addlegendimage{/pgfplots/refstyle=businesses}\addlegendentry{Restaurants}

            \addplot +[draw=blue]
            table[x expr=\coordindex,y=reviews] {\datatabletwo};
            \label{restreviews}

            %\addlegendimage{/pgfplots/refstyle=reviews}
            \addlegendentry{Reviews}
        \end{axis}
    \end{tikzpicture}
    \end{subfigure}
    \caption{Businesses (left) and Restaurants (right) distribution together with the number of reviews associated with each category.}
    \label{fig:businesses-distribution}
\end{figure}

\subsubsection{Modeling \textit{Yelp.com} using hypergraphs.}\label{hyelp}
\begin{figure}[t!]%hb!
    \centering
  \includegraphics[width=.27\linewidth]{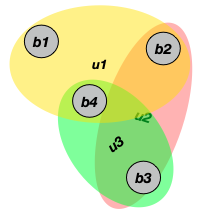}
  \caption{The Yelp hypergraph defined by user reviews.}
  \label{fig:hreviews}
\end{figure}

We modeled the \textit{Yelp} dataset using a hypergraph $H=(V,E)$, where $V$ represents Yelp businesses, and $E$ represents Yelp users. In more detail, each hyperedge representing a user $u$ contains all businesses $u$ has written at least one review for.
Figure~\ref{fig:hreviews} shows a simple example hypergraph, defined by four businesses ($V=\{b_1,b_2,b_3,b_4\}$) and three users ($E=\{u_1,u_2,u_3\}$). Here, the hyperedge $u_1$ connects the businesses $b_1, b_2,$ and $b_4$, as the corresponding user has written at least one review for each of the listed business.

%Figure~\ref{fig:hreviews} shows a simple hypergraph example of such data. As shown in the figure, the hypergraph $H$ is defined by four businesses ($V=\{b_1,b_2,b_3,b_4\}$) and three users ($E=\{u_1,u_2,u_3\}$).
%Hyperedges correspond to three different users that have written a review for a certain business.
%For instance, hyperedge $u_1$ connects businesses $b_1, b_2,$ and $b_4$, as the corresponding user have written reviews for each of the listed business.

%Since processing the entire Yelp dataset is a challenging computational task, to accomplish our analysis, we decided to explore only a subset of the data set. 
To accomplish our analysis, we explored only a subset of the Yelp data set, given the massive amount of available data (around 9 GB). 
We modeled the Yelp hypergraph according to the two following strategies.
\begin{enumerate}
    \item \textbf{yelp-dataset1.}
    We collected $1$ million of randomly chosen reviews, from which we selected the businesses and the users to build the hypergraph. It is worth mentioning that such selection defined the total number of businesses and users involved, i.e., the size of the hypergraph itself. We run our analysis on the largest connected component of the so-built hypergraph, removing isolated vertices and small components. More detailed information about the dimension of the network can be found in the following section.
    %We randomly selected a subset of reviews, which consequently defined the number of businesses considered for the analysis.
    % of a fixed size
    %It is worth mentioning that such selection also defines the number of businesses involved. Our analysis has been run on connected hypergraphs, obtained by removing isolated vertices and small components, from a collection of $1$ million of reviews uniformly at random chosen.

    %\todo{to check}
    \item \textbf{yelp-dataset2.} 
    To generate this data set, we focused our attention only on the businesses belonging to the category ``restaurant''. We attached to each restaurant the label (selected from its categories set) representing the type of cuisine it offers according to the methodology described above.
    %As some restaurants may offer different types of cuisine, represented as Yelp sub-categories, we selected one category from its categories set according to the frequencies (highest) in the whole dataset. 
    %The computed dataset generates five hypergraphs described in Table~\ref{table_graphs_par}.
    %(some businesses may have more than one category; in such cases, we select one category from its categories set according to the frequencies (highest) in the whole dataset). %\todo{PP: I hope I got this right. Please check. }
    % is a businesses subset defined by only the restaurants' category (the category with higher frequency in the dataset), each business restaurant can have more than one sub-category, we assign to a restaurant only one category according to their frequency value in the data.
\end{enumerate}

%\subsubsection{Results}

\subsubsection{Forecasting the number of stars for a new business.} 
This experiment focuses on the forecasting of the number of stars of a given business $v$, based on the information available in the local neighborhood of $v$. We developed two different strategies. We based one approach on the information provided by the hypergraph $H$ defined in the previous section, the other on the information provided by the corresponding weighted two-section graph. Here, the weight of an edge ($u,v$) corresponds to the number of users that reviewed both $u$ and $v$; namely, the number of hyperedges containing both $u$ and $v$.

In the \textit{hypergraph-based} strategy, for each business $u$, we first computed the average number of stars for all hyperedges containing $u$ (where, in each hyperedge $e$, the average was computed excluding $u$). This estimated value corresponded to the average rating given by the user associated with $e$. Then, we obtained the prediction about the number of stars of $u$ as the average over the values computed in the previous step.
In other words, the forecast of the number of stars of a given user $u$ is the average over the averages in each hyperedge involving $u$.
Formally,
\begin{eqnarray*}
s_i'(u) = \dfrac{1}{|E(u)|} \sum\limits_{e\in E(u)} \left( \dfrac{1}{|e|-1} \sum\limits_{v\in e, v\neq u}   s(v)  \right),
\end{eqnarray*}
where $s(v)$ denotes the number of stars associated to $v$, $E(v)$ denotes the set of hyperedges that contains $v$, and $s_i'(u)$ denotes the forecasted value for $u$ for strategy $i$.

The \textit{graph-based} strategy exploits the weighted two-section graph of $H$. In this case, the forecast of the number of stars of a user $u$ is the weighted average over the neighborhood of $u$.
Formally,
\begin{eqnarray*}
s_2'(u)= \dfrac{\sum\limits_{e=(u,v)\in E}  s(v) w(e)  } { \sum\limits_{e=(u,v)\in E}   w(e)},
\end{eqnarray*}
where $w(e)$ denotes the weight of edge $e$.

To compare these two strategies, we computed the average error as follows:
\begin{eqnarray*}
err_i = \dfrac{\sum\limits_{u \in V} |s(u)-s_i'(u)|}{|V|}.
\end{eqnarray*}

We performed this experiment on several instances of the \textit{yelp-dataset1}, varying the number of considered reviews. In particular, we selected five subsets of increasing size equal to $250$k, $500$k, $750$k, and $1$ million. For each subset, we computed the corresponding hypergraph. All hypergraphs resulted in having 209,393 nodes, while the number of hyperedges ranged from 175,022 to 432,381.

\begin{figure}[ht!]
  \begin{center}
    \begin{tikzpicture}[scale=1.0]
      \begin{axis}[
          width=.5\linewidth, % Scale the plot to \linewidth
          grid=major,
          grid style={dashed,gray!30},
          xlabel= $|Reviews|$,
          ylabel=$err$,
          ylabel style={rotate=-90},
          legend pos=south east,
          ymin=0, ymax=1.0, xmin=200000, xmax=1000000, scaled x ticks=false, xtick distance=250000
        ]
        \addplot
        table[x=V,y=G,col sep=comma] {task1.csv};
           \addplot
        table[x=V,y=H,col sep=comma] {task1.csv};
        \legend{Graph,Hypergraph}
      \end{axis}
    \end{tikzpicture}
    % \begin{tikzpicture}[scale=0.90]
    %   \begin{axis}[
    %       width=.5\linewidth, % Scale the plot to \linewidth
    %       grid=major,
    %       grid style={dashed,gray!30},
    %       xlabel= $|Reviews|$,
    %       ylabel=,
    %       ylabel style={rotate=-90},
    %       legend pos=south east,
    %       ymin=0, ymax=1.0, xmin=200000, xmax=1000000, scaled x ticks=false, xtick distance=250000
    %     ]
    %     \addplot
    %     table[x=V,y=G,col sep=comma] {task1restaurants.csv};
    %       \addplot
    %     table[x=V,y=H,col sep=comma] {task1restaurants.csv};
    %     \legend{G, H}
    %   \end{axis}
    % \end{tikzpicture}
    \caption{Average error of the stars’ forecast experiment performed on yelp-dataset1, varying the number of reviews used to build the hypergraph and the corresponding two-section graph.\label{fig:taskone}}
  \end{center}
\end{figure}
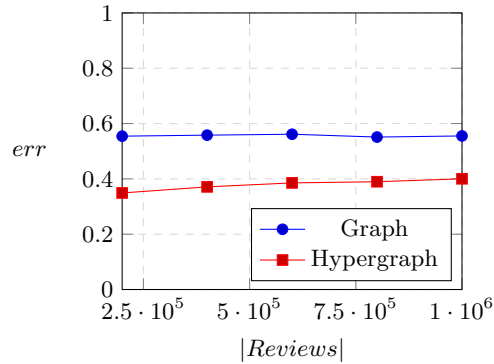

Figure~\ref{fig:taskone} depicts the results for this experiment. As clearly visible from the plot, the error value $err_2$ using the weighted two-section graph is always greater than the error value $err_1$ obtained exploiting the hypergraph representation.
We also run the stars forecasting experiment on the \textit{yelp-dataset2}, which generated an hypergraph with 35,466 nodes and 1,133,890 hyperedges. Also in this case, the error for the two-section graph and the hypergraph was always close to $0.6$ and $0.5$, respectively. 
%We obtained similar results, where the error for the two-section graph and the hypergraph was always close to $0.6$ and $0.5$, respectively. 
%was always close to $0.6$ while the error for the hypergraph representation is always close to $0.5$. 
%(35466, 1133890)
%
%Both experiment instances highlighted promising results: as the average error obtained by the business stars prediction task was around $0.5$, it is crucial to be able to predict low rated instances accurately. 
Interestingly, both experiment instances suggest that the information provided by a hypergraph model is more accurate than the information provided by the corresponding weighted two-section graph. 

\subsubsection{How much the rating of a review burden on a business?} \label{par:yelp_exp2} 
This second experiment examines the amount of information conveyed by different kinds of reviews, according to the number of stars associated with them. We used the \textbf{yelp-dataset2}; due to performance issues, we restricted the set of businesses only to the restaurant category, as described in Section~\ref{hyelp}. %We built five hypergraphs after having partitioned the reviews into five categories: $1$ stars, $2$ stars, \ldots, $5$ stars.
We ended up with $342{,}044$ $\textbf{1}$-star reviews, $281{,}307$ $\textbf{2}$-star reviews, $402{,}053$ $\textbf{3}$-star reviews, $791{,}068$ $\textbf{4}$-star reviews, and  $1{,}188{,}558$ $\textbf{5}$-star reviews. We built five hypergraphs, one for each set of reviews. Henceforth, we will denote with $H_i$, for $i=1,2,\ldots,5$, the hypergraph generated using the set of reviews having $i$ stars and by $G_i$ the corresponding two-section graph.
We first computed some statistics on these five hypergraphs and their corresponding two-section graphs. The collected information can be found in Table~\ref{table_graphs_par}. 
This preliminary analysis highlighted that the so-built hypergraphs were quite different from their graph counterpart. 
Interestingly, what stands out in the columns corresponding to the two-section graphs is that the number of both edges and triangles exhibits a ``bell-shaped'' trend as a function of the number of stars.

\begin{table}[b!]

\centering
\begin{tabular}{cccrr}
%\hline
\toprule
\textbf{Stars} & \textbf{$H_i$ $(|V|;|E|)$} & $G_i$ $(|V|;|E|)$ & $G_i$ Modularity & $G_i$ Triangles \\ 
\midrule
1 & (29479; 244671)	&	(29479; 240412)	&	0.6210	&	1,158,341 \\ %\hline
2 & (28055; 173140)	&	(28055; 484527)	&	0.7173	&	6,491,497 \\ %\hline
3 & (30369; 177792)	&	(30369; 2636712)	&	0.6616	&	289,584,451 \\ %\hline
4 & (32987; 301578)	&	(32987; 4384044)	&	0.6857	&	404,709,664 \\ %\hline
5 & (32558; 590320)	&	(32558; 2187473)	&	0.6657	&	104,128,714 \\ %\hline
\bottomrule
\vspace{5pt}
\end{tabular}
\caption{Some statistics on the yelp-dataset2 dataset.\label{table_graphs_par} }
\end{table}

In this second experiment, we focused our attention on understanding which factors influence the chance that a given user reviews any two businesses. To this aim, we tested the ability of these two different models to detect the community structure of the network. In other words, we intended to find the division of the vertices set into groups of restaurants that were similar among themselves but dissimilar from the rest of the network, and getting insights about the nature of their similarity.
To have a first insight about the nature of the so-built network, we manually partitioned the five hypergraphs $H_i$, for $i=1,2,\ldots,5$, according to several restaurant properties available in the Yelp dataset. In particular, we considered the following attributes: the location of the restaurant (\textit{city}, and \textit{state}), whether it sells \textit{alcohol}, its \textit{noise level}, whether it offers a \textit{take away} service, and its food \textit{category}.
Table~\ref{table_hypgraph} contains the modularity values for the $6$ different partitions evaluated. 
To calculate the modularity, we used the approach presented in~\cite{kaminski2018clustering} and implemented in SimpleHypergraphs.jl (see Section ~\ref{ssec:analytical}). As shown in the table, the modularity is strongest when we used geographical information, as cities or states, to partition the hypergraph. These higher values mean that i) people doing reviews usually visit restaurants within the same city and that ii) if restaurants in different cities are reviewed by a single person, they are usually in the same state. It is worth noting that $1$-star reviews have the strongest modularity values across all partitions. This result suggests that there is a group of people who have a stronger tendency to submit negative scores on the base of some ground-truth property of a restaurant.

\begin{table}[t!]
\centering
\setlength{\tabcolsep}{2pt}
\begin{tabular}{cccccccc}
\toprule
\textbf{Stars} & $H_i$ $(|V|;|E|)$  & \textbf{City}   & \textbf{State}  & \textbf{Alcohol} & \textbf{Noise Level} & \textbf{Take Ou}t & \textbf{Category} \\ %\hline\hline
\midrule
1 & (29479; 244671)	&	0.8833	&	0.9562	&	0.8166	&	0.8104	&	0.8176	&	0.8163 \\ %\hline
2 & (28055; 173140) 	&	0.8582	&	0.9462	&	0.7744	&	0.7651	&	0.7731	&	0.7702 \\ %\hline
3 & (30369; 177792) 	&	0.8132	&	0.9226	&	0.7075	&	0.6940	&	0.6966	&	0.6965 \\ %\hline
4 & (32987; 301578) 	&	0.7812	&	0.9081	&	0.6573	&	0.6385	&	0.6419	&	0.6400 \\ %\hline
5 & (32558; 590320) 	&	0.8027	&	0.9145	&	0.6963	&	0.6797	&	0.6894	&	0.6841 \\ %\hline
\textbf{ALL} & (35856; 950488)	&	0.7500	&	0.8985	&	0.6162	&	0.5919	&	0.6013	&	0.5967 \\ %\hline
\bottomrule
\vspace{5pt}
\end{tabular}
\caption{Hypergraphs size for each $H_i$, $i=1,2,\ldots,5$. The modularity values have been computed on $6$ different manually-evaluated ground-truth partitions, conditioned on some properties of the Yelp restaurants. Experiment have been run on yelp-dataset2.\label{table_hypgraph} }
\end{table}

Based on these outcomes, we then investigated to what extent the partition obtained from a community detection algorithm was able to mimic the communities output of a ground-truth partitioning.  We compared the communities obtained on both models (hypergraph and two-section graph) against the given ground-truth partition to catch the best model in capturing that specific feature of the underlying network. Specifically, we focused on the communities evaluated on the ``type of cuisine'' (food \textit{category}) of each restaurant. The ground-truth partition was made up of $55$ categories, of which the largest (American Traditional) comprised $7{,}107$ restaurants.

Several community detection algorithms have been proposed in the literature. A review of the various methods available can be found, for example, in~\cite{Danon05comparingcommunity,10.1007/978-3-319-27308-2_30}. We opted for the label propagation (LP) strategy proposed by Raghavan et al.~\cite{Raghavan2007NearLT} to find communities in a graph. It can be summarized as follows. Initially, each node has a unique label (initialization phase). At each iteration step, each node's label is updated by choosing the most frequent label in its neighbors (propagation rule); ties are broken with a random choice. The algorithm terminates either if it does not modify any label in two consecutive iterations, or it hits the predefined number of iterations (termination criteria). We exploited the LP implementation provided by the Julia \texttt{LightGraphs} library~\cite{LightGraphs}.
For hypergraphs, we propose to define a new LP strategy which generalizes the algorithm in~\cite{Raghavan2007NearLT}.
The proposed algorithm shares both the initialization phase as well as the termination criteria with the standard graph label propagation algorithm. We modified the propagation rule, splitting it into two phases: hyperedge labeling and vertex labeling. 
During the hyperedge labeling phase, the labels of the hyperedges are updated according to the most frequent label among the vertices contained in that hyperedge. Similarly, during the vertex labeling phase, the label of each vertex is updated by choosing the most frequent label among the hyperedges it belongs to.
Both algorithms have been executed setting the maximum number of iterations to $100$.

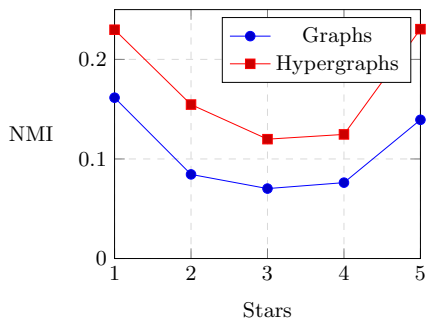
\begin{figure}[h!]
  \begin{center}
    \begin{tikzpicture}[scale=0.9]
      \begin{axis}[
          width=.5\linewidth, % Scale the plot to \linewidth
          grid=major,
          grid style={dashed,gray!30},
          xlabel= Stars,
          ylabel=NMI,
          ylabel style={rotate=-90},
          legend pos=north east,
          ymin=0.0, ymax=0.25, xmin=1, xmax=5, scaled x ticks=false, xtick distance=1
        ]
        \addplot
        table[x=NMI,y=G,col sep=comma] {nmi.csv};
           \addplot
        table[x=NMI,y=H,col sep=comma] {nmi.csv};
        \legend{Graphs,Hypergraphs}
      \end{axis}
    \end{tikzpicture}\vspace*{-0.2truecm}
    \caption{NMI values between the ``type of cuisine'' ground-truth partition and the $10$ partitions obtained running the label propagation algorithm on the five hypergraphs and on the corresponding 2-section graphs. Experiment have been run on yelp-dataset2.  \label{fig:NMI}}
  \end{center}
\end{figure}

To evaluate the correlation between two partitions, several measures have been borrowed from information theory, for instance, the \textit{Normalized Mutual Information} (NMI) coefficient, which considers each partition as a probability distribution. Several variants of the  NMI  have been introduced (see, for example,~\cite{Vinh:2010:ITM:1756006.1953024} for a detailed discussion). In this paper, we use the \textit{sum} variant, which is defined as follows:
\begin{equation}
  NMI(X,Y)=\frac{I(X,Y)}{H(X)+H(Y)},
\end{equation}
where $I(X,Y)$ denotes the \textit{Mutual Information} (i.e., the shared information between the two distributions $X$ and $Y$)
and $H(X)$ denotes the Shannon Entropy (i.e., the information contained in the distribution) of $X$.
The NMI coefficient holds several interesting properties. One of the most useful is that it is a \textit{metric}, and it lies within a fixed range $[0,1]$. Specifically, it equals $1$ if the partitions are identical, whereas it has an expected value of $0$ if the two partitions are independent.
Results appear in Figure~\ref{fig:NMI}.
Although the correlation, in general, is not very high (the best result is $0.23$ for $H_5$), the figure provides two interesting points. First, in all five cases, the quality of partitioning provided by hypergraphs is always better than that provided by the corresponding two-section graphs.
Moreover, also in this second experiment, results appear in the form of an ``inverted bell shape'' (the best results, in this case, are given by the two external values). In a sense, very good as well as very bad reviews are much better able to identify restaurants genre.

% \begin{figure}[ht!]
%     \centering
%   \includegraphics[width=0.65\linewidth]{NMI.png}
%   \caption{Then NMI between the ground truth partition and the $10$ partitions obtained running the label propagation algorithm on the five hypergraphs and on the corresponding 2-section views.}
%   %improve label
%   \label{fig:NMI}
% \end{figure}

\subsubsection{Performance discussion.}

We performed our experiments on a Linux Ubuntu $18.04$ machine equipped with an Intel i$7$ processor endowed of
$8$ cores, $16$ GB of memory, and $256$ GB SDD  disk. We implemented the experiments using the $1.3.1$ Julia language version. In the following, we present the average performance, in seconds, obtained executing $10$ runs for each experiment. 
With this configuration, Julia required about $117.68$ seconds to load the Yelp data (about $9$GB) in memory.

The first case study concerned the execution of the forecasting algorithm on each hypergraph and the corresponding two-section graph. The completion time for the biggest experiment was about $42$ seconds, while the smallest one required $0.92$ seconds. In the second use case, the average completion time for all the experiments was around $20$ seconds. Specifically, for all subsets of \textit{yelp-dataset2} (see Section 4), we computed the network communities - exploiting the LP algorithm - on both the hypergraphs and the corresponding two-section graphs. Moreover, we compared the goodness of the computed sets with a ground-truth partition evaluating the NMI coefficient.

\subsection{Mining and modeling social relationships: \textit{Game of Thrones}}
\label{case2}

%Another interesting line of inquiry is grasping the intricacies of a literary work or a movie to get insights into the narrative structure finding common patterns across several plot lines~\cite{bonato2016}, make sense of intricate character relationships~\cite{beveridge2018got}, or just to have fun trying to predict how the plot itself will evolve~\cite{stavanja2019predicting}. 
Another interesting line of inquiry is grasping the intricacies of a literary work or a movie to get insights into their narrative structure. Various works focused on finding common patterns across several plot lines~\cite{bonato2016}, making sense of intricate character relationships~\cite{beveridge2018got}, or just having fun trying to predict how the plot itself will evolve~\cite{stavanja2019predicting}. 
Usually, the character interaction network is modeled with a graph, where a vertex represents a storyline character, and an edge points out an interaction between two characters. Edges may also have different nature; for instance, they can express that the names (or aliases) of two characters appear within a certain number of words apart~\cite{agarwal2012alice,bonato2016}, that a character $A$ has spoken right after a character $B$, or that a character $A$ and character $B$ appear in a scene together~\cite{beveridge2018got}. 
The output graph is, commonly, an indirect and weighted~network.

A network built this way is an example of artificial collaboration networks, as usually most pairs of characters have cooperated or have been antagonist one of another~\cite{alberich2002marvel}.
A typical analysis carried on this kind of network is determining the most important characters, according to several centrality measures. Some nodes play a massive role in the network, either by having many connections or by being strategically positioned to help connect distant parts of the network. A character may, indeed, be relevant or influential with different facets, and it is fundamental to interpret these quantities with respect to the underlying domain.
Another question an exploratory analysis of networks of characters aims to answer is to capture which characters naturally belong together, forming coherent communities within the network. Investigating these behavioral patterns means looking for coherent sub-plotlines hidden in the network~\cite{networkofthrones}.

Considering a character interactions network that is built upon characters' co-occurrence in movie scenes, it is straightforward to see that this kind of network can be naturally modeled with a hypergraph, where vertices are still associated with characters and hyperedges are related to scenes. In this case, the topology itself of the hypergraph allows us to easily find clusters of characters that commonly appear together within a movie or a TV series episode or season. 
To verify if hypergraphs are able to convey more information than a standard graph analysis approach also in the case of collaboration networks, we have modeled and analyzed the Game of Thrones TV series characters co-occurrence. 
As for the Yelp use case (see Section~\ref{sec4}), we compare the findings obtained by exploiting hypergraphs with the results achieved using the two-section graphs. 
%In this experiment, we are interested in exploring the structure of the interactions between characters, which appear in the same season, we are able to understand the characters' collaborations. Moreover, we look for the roles (or importance) of characters along the seasons.  

%results
%communities
%centrality measures

%\newpage
\subsubsection{The Game of Thrones TV series dataset.}
Game of Thrones~\cite{gameofthrones} (GoT) is an American fantasy drama TV series, created by D.\ Benioff and D.B.\ Weiss for the American television network HBO. It is the screen adaption of the series of fantasy novels \textit{A Song of Ice and Fire}, written by George R.R.\ Martin. The series premiered on HBO in the United States on April 17, 2011, and concluded on May 19, 2019, with 73 episodes broadcast over eight seasons. With its 12 million viewers during season 8 and a plethora of awards---according to Wikipedia~\footnote{\url{https://en.wikipedia.org/wiki/Game\_of\_Thrones}}---Game of Thrones has attracted record viewership on HBO and has a broad, active, and international fan base. 
The intricate world narrated by George R.R.\ Martin and scripted by Benioff and Weiss extend well beyond the boundaries of the traditional TV medium to create a deeply immersive entertainment experience~\cite{media_gameofthrones}. This allows both the academic community and industries to study not only complex dynamics within the GoT storyline~\cite{beveridge2018got}, but also how viewers engage with the GoT world on social media~\cite{antelmi2018got,perez2019got,Rhodes2019got}, or how the novel itself is a portrait of real-world dynamics~\cite{lovric2019got,olesker2019got,milkoreit2019got,Muno2019got,zare2019got}.

This study is based on the dataset at the GitHub repository \textit{Game of Thrones Datasets and Visualizations}\footnote{Game of Thrones Datasets and Visualizations \url{https://github.com/jeffreylancaster/game-of-thrones} by Jeffrey Lancaster}. Specifically, we made use of the data describing season episodes and containing meta-information about each of them, such as title, identification number, season, and description. Information about each scene within an episode is also reported. For each scene, start, end, location and a list of characters performing in it are listed.
Table~\ref{tab:tbl-got-fields} reports some basic information about the number of episodes, scenes, and characters per GoT season.
A more detailed description of the dataset can be found on the GitHub repository.

\begin{table}[ht]

    \centering
    %\begin{tabular}{lcp{0.7\textwidth}}
    \begin{tabular}{>{\centering}p{0.15\textwidth}>{\centering}p{0.15\textwidth}>{\centering}p{0.15\textwidth}p{0.15\textwidth}<{\centering}}
        \toprule
            \textbf{Season} & \textbf{Episodes} & \textbf{Scenes} & \textbf{Characters} \\
            \midrule
            1 & 10 & 286 & 125 \\
            2 & 10 & 468 & 137 \\
            3 & 10 & 470 & 137 \\
            4 & 10 & 517 & 152 \\
            5 & 10 & 508 & 175 \\
            6 & 10 & 577 & 208 \\
            7 & 7 & 468 & 75 \\
            8 & 6 & 871 & 66 \\
        \bottomrule
        \vspace{5pt}
    \end{tabular}
    %\vspace{0.2cm}
    \caption{Some GoT dataset numbers. \label{tab:tbl-got-fields}}
   
\end{table}

\subsubsection{Modeling Game of Thrones using hypergraphs.}
\begin{figure}[!htbp]
  \centering
  \includegraphics[width=.45\columnwidth]{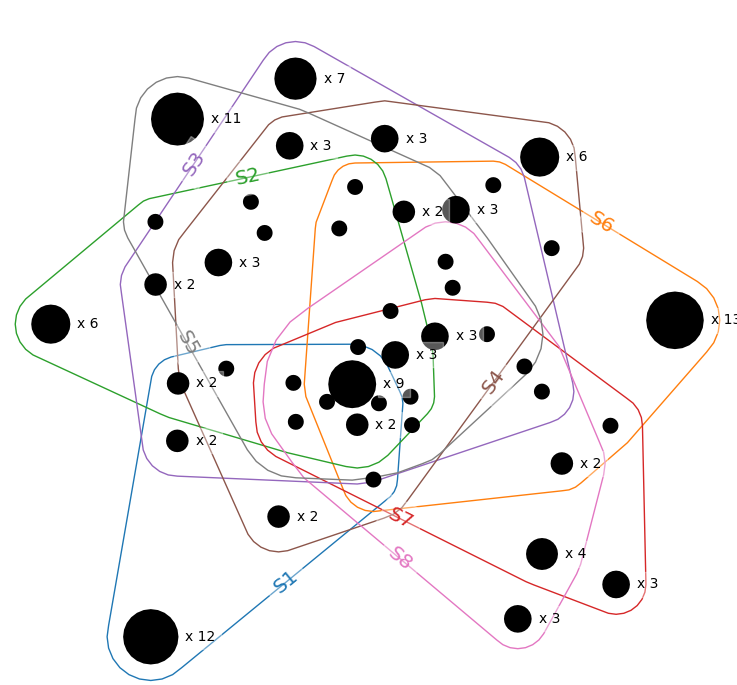}
  \caption{The GoT season hypergraph  $H_{seasons}$ defined by characters overlapping through seasons.}
  \label{fig:hgot}
\end{figure}

\begin{figure}[!htbp]
%\begin{sidewaysfigure*}[ht]
    \centering
    \begin{subfigure}{.47\columnwidth}
      \centering
      % include first image
      \includegraphics[width=\linewidth]{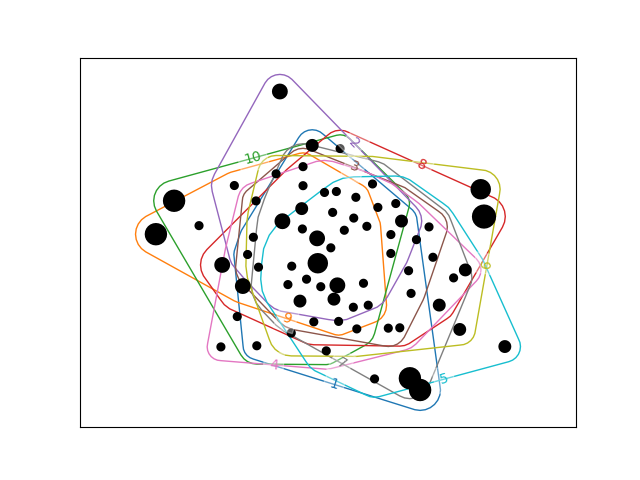} \vspace*{-0.8truecm}  
      \caption{$H^{S1}_{episodes}$}
      \label{fig:sub-first}
    \end{subfigure}
    \begin{subfigure}{.47\columnwidth}
      \centering
      % include second image
      \includegraphics[width=\linewidth]{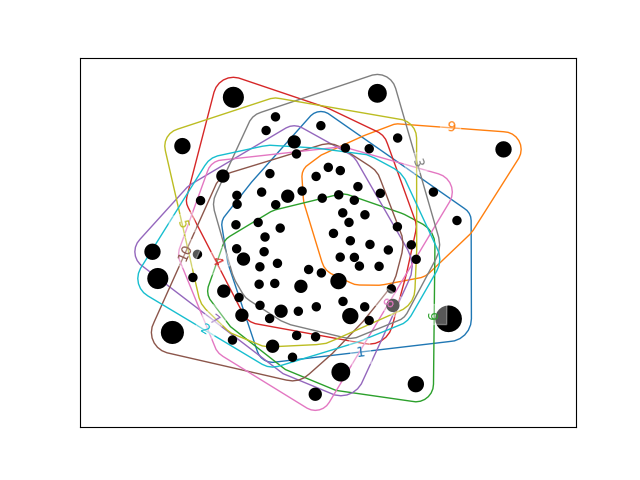} \vspace*{-0.8truecm}
        \caption{$H^{S2}_{episodes}$} 
      \label{fig:sub-second}
    \end{subfigure}
    
    \centering
    \begin{subfigure}{.47\columnwidth}
      \centering
      % include first image
      \includegraphics[width=\linewidth]{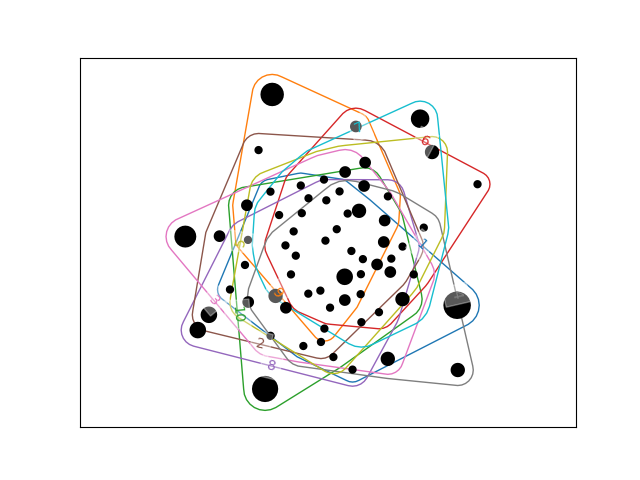}  \vspace*{-0.8truecm}  
      \caption{$H^{S3}_{episodes}$}
      \label{fig:sub-first}
    \end{subfigure}
    \begin{subfigure}{.47\columnwidth}
      \centering
      % include second image
      \includegraphics[width=\linewidth]{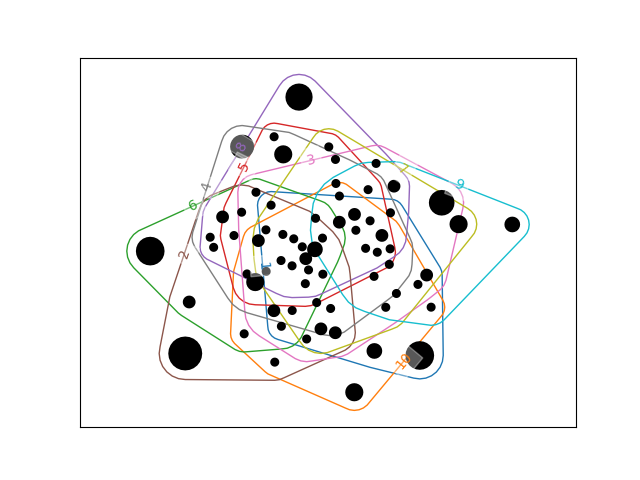}  \vspace*{-0.8truecm}  
       \caption{ $H^{S4}_{episodes}$}
      \label{fig:sub-second}
    \end{subfigure}

        \centering
    \begin{subfigure}{.47\columnwidth}
      \centering
      % include first image
      \includegraphics[width=\linewidth]{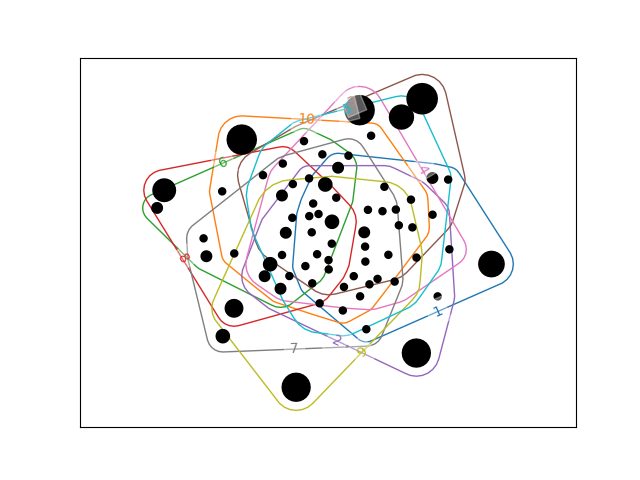}  \vspace*{-0.8truecm}  
      \caption{$H^{S5}_{episodes}$}
      \label{fig:sub-first}
    \end{subfigure}
    \begin{subfigure}{.47\columnwidth}
      \centering
      % include second image
      \includegraphics[width=\linewidth]{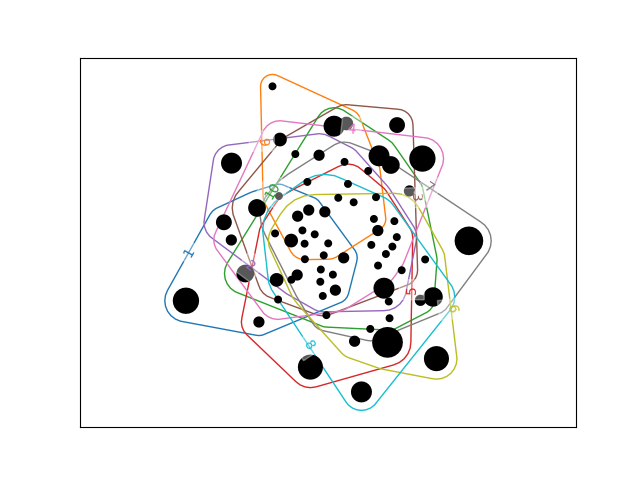}  \vspace*{-0.8truecm}  
  \caption{$H^{S6}_{episodes}$}
      \label{fig:sub-second}
    \end{subfigure}

        \centering
    \begin{subfigure}{.47\columnwidth}
      \centering
      % include first image
      \includegraphics[width=\linewidth]{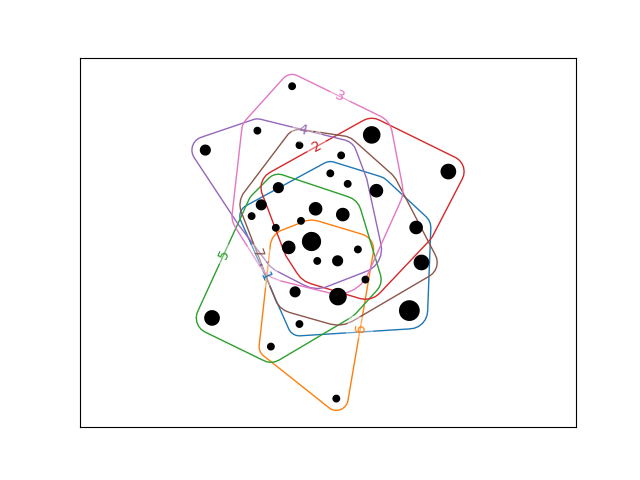}  \vspace*{-0.8truecm}  
     \caption{$H^{S7}_{episodes}$}
      \label{fig:sub-first}
    \end{subfigure}
    \begin{subfigure}{.47\columnwidth}
      \centering
      % include second image
      \includegraphics[width=\linewidth]{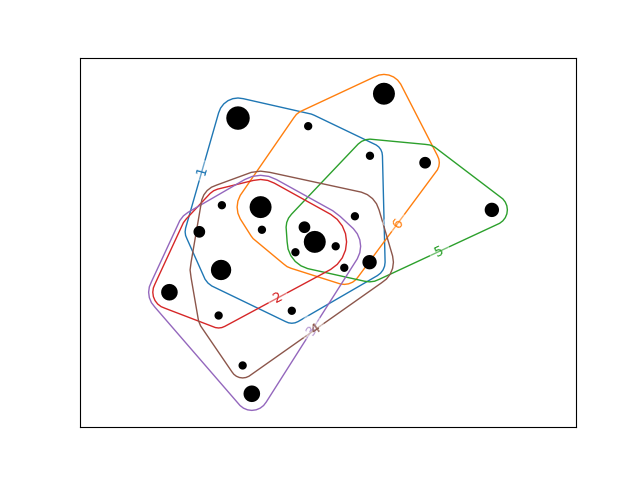}  \vspace*{-0.8truecm}  
     \caption{$H^{S8}_{episodes}$}
      \label{fig:sub-second}
    \end{subfigure}
\caption{GoT characters overlap through Episodes$\times$Season.} 
\label{fig:got_dc}
\end{figure}

We studied GoT characters' co-occurrences with different levels of granularity. We modeled the GoT dataset building three different types of hypergraphs, each one reporting whether the GoT characters have appeared in the same season, in the same episode, or in the same scene together. In the following, we describe the hypergraphs considered in our study:
\begin{itemize}
    \item \textbf{Seasons.}
        A coarse--grained model is represented by the seasons hypergraph $H_{seasons} = (V, E_{seasons})$, where $V$ represents GoT characters, and $E_{seasons}$ represents, for each GoT season $s$, the set of characters appearing in $s$. The hypergraph $H_{seasons}$ is shown in Figure~\ref{fig:hgot}, using an Euler-based visualization (see Section~\ref{ssec:hg_viz}). 
    
    \item \textbf{Episodes$\times$Season.}
        An intermediate-grained model is obtained by considering co-occurrence within Episodes.
        In this case, we considered $8$ different hypergraphs, $H^s_{episodes} = (V^s, E^s_{episodes}),$ $  s \in [1, 8]$, where $s$ indicates a GoT season, $V^s$ represents the GoT characters appearing in season $s$, and $E^s_{episodes}$ represents, for each GoT episode $e$ of season $s$, the set of characters appearing in $e$.
        The $8$ hypergraphs  $H^s_{episodes}$ are shown in Figure~\ref{fig:got_dc}. 
        
    \item \textbf{Scenes$\times$Season.}
        A fine-grained model is obtained by considering co-occurrence within scenes.
        In this case, we considered $8$ different hypergraphs, $H^s_{scenes} = (V^s, E^s_{scenes}), s \in [1, 8]$,  where $s$ indicates a GoT season, $V^s$ represents the GoT characters appearing in season $s$, and $E^s_{scenes}$ represents, for each GoT scene $f$ of season $s$, the set of characters appearing in $f$.
    \end{itemize}

%basic stats for each one + twosection view

%data about hg/g per season

\subsubsection{The collaboration structure of Game of Thrones.} 
We performed a community detection task on the Scenes$\times$Season hypergraphs and their corresponding two-section graphs. Running the community detection algorithm on such networks allows us to find coherent plotlines and, therefore, groups of characters frequently appearing in a scene together. In this experiment, our goal is to figure out whether and to what extent hypergraphs are able to capture characters' collaboration (and, generally speaking, any type of user-defined collaboration) with respect to graphs.
Specifically, we executed the label propagation algorithms defined in Section~\ref{par:yelp_exp2} and measured the quality of the solution obtained by computing the \emph{modularity}. 
Results, described in Table~\ref{tab:communities}, show that the solutions obtained for hypergraphs provide a higher number of communities. This pattern emerges particularly in the graphs describing the last two seasons, which are characterized by a smaller number of characters. 

To measure the difference between the two approaches, we computed the NMI between the obtained partitions, shown in Table~\ref{tab:nmi}. Considering the  NMI values, the partitions are strongly related,  except for the last two seasons (the seasons with the lower number of characters). It is important to notice that the last two seasons exhibit the worst results. Nonetheless, we believe that it is well justified to the nature of their screenplay: fewer characters and high related plotlines. The equivalent hypergraphs are characterized by a high degree for both nodes and hyperedges. As a consequence, the corresponding two-section views result in \textit{quasi}-complete graphs. In this particular case, the label propagation algorithm is not able to find out distinct communities. 

\begin{table}[h!]

\centering
\begin{adjustbox}{max width=\textwidth}
\begin{tabular}{c c c c c c c c c}
\toprule
           &\textbf{$H^{S1}_{scenes}$} & \textbf{$H^{S2}_{scenes}$} &\textbf{$H^{S3}_{scenes}$} & \textbf{$H^{S4}_{scenes}$}&\textbf{$H^{S5}_{scenes}$} & \textbf{$H^{S6}_{scenes}$} &\textbf{$H^{S7}_{scenes}$} & \textbf{$H^{S8}_{scenes}$}\\
           &$(|C|, m)$&$(|C|, m)$&$(|C|, m)$&$(|C|, m)$&$(|C|, m)$&$(|C|, m)$&$(|C|, m)$&$(|C|, m)$  \\
\midrule
\textbf{\textit{H}} \hspace{5pt}& $(12, .5359)$   & $(15, .4511)$   & $(15, .7028)$   & $(18, .5527)$  & $(22, .5794)$  & $(19, .5781)$   & $(11, .2159 )$  & $(8, .1536)$    \\
\textbf{\textit{G}}  \hspace{5pt}& $(8, .3399)$  & $(9, .6970)$  & $(11, .7652)$  & $(11, .6218)$   & $(15, .7300)$   & $(16, .7342)$  & $(4, 2163)$   & $(1, .0)$\\
\bottomrule
\vspace{5pt}
\end{tabular}
\end{adjustbox}
\caption{A comparison between hypergraph and graph capability on discovering  characters' communities.
For each hypergraph ($H$) and its two-section graph ($G$), the table provides the number of communities ($|C|$) and the corresponding modularity values ($m$). \label{tab:communities}}

\end{table}

\begin{table}[t!]

\centering
\begin{tabular}{ccccccccc}
\toprule
           &\textbf{$S1$} & \textbf{$S2$} &\textbf{$S3$} & \textbf{$S4$}&\textbf{$S5$} & \textbf{$S6$} &\textbf{$S7$} & \textbf{$S8$}\\
\midrule
$NMI$ & $.82085$   & $.83354$   & $.92340$   & $.87596$  & $.88143$  & $.88143$   & $.69506$  & $.0$ \\
\bottomrule
\vspace{5pt}
\end{tabular}
\caption{NMI values evaluated between the partitions obtained computing the communities on the Scenes$\times$Season hypergraphs and their corresponding two-section graphs. \label{tab:nmi}}
%\vspace{0.2cm}
\end{table}

\subsubsection{Discussion on season 8.} 
It is worth discussing the interesting facts revealed by the results of the 8th season. The label propagation algorithm, in the case of the two-section graph, reveals only one big community --- the whole graph itself. However, in the case of hypergraphs, it can determine eight different communities. In the following, we discuss the conflicting results obtained by providing a possible interpretation for the $8$ discovered communities by the label propagation algorithm run on the hypergraph model.

In more detail, three minor communities of characters, appearing only in few scenes in the whole season, emerged from the (hyper)network structure. These communities are made up by: i) the \textit{Winterfell} boy --- appearing only in the first episode; ii) \textit{Dirah}, \textit{Craya}, \textit{Marei} --- seen only in the first episode; and iii) \textit{Eleanor} and her daughter --- occurring only in the fifth episode. Exploiting a hypergraph approach, the algorithm correctly identifies background characters that do not contribute to the main storyline and that are not strictly related to any main character.
Other two communities pinpoint key characters belonging to the two central alliances: the \textit{north} versus the \textit{south}. The south-alliance is made up of \textit{Cercei Lannister}, her counselor \textit{Qyburn}, her guard \textit{Gregor Clegane}, and her husband \textit{Euron Greyjoy}; while the north-alliance is forged by Jon Snow and \textit{Daenerys Targaryen}, with her dragons. In particular, in the north-alliance community also appear two enemies that have been faced by \textit{Jon} and \textit{Daenerys}: the \textit{Night King} (and his white walker soldiers), and Harry Strickland, captain of the sellsword \textit{Golden Company}.
Two more communities can be labeled as north-allied: they contain a group of characters that consistently have interacted and/or fought together. Indeed, one group contains \textit{Bran Stark} and \textit{Theon Greyjoy} (who stands guard for him  in the battle for Winterfell), Lyanna Mormont and Wun Wun (they fought against in the battle for \textit{Winterfell}), \textit{Lord Varys}, \textit{Davos Seaworth}, \textit{Grey Worm} and \textit{Jorah Mormont}. The other community includes \textit{Arya} and \textit{Sansa Stark}, \textit{Samwell Tarly} and his partner \textit{Gilly}, \textit{Brienne of Tarth} and \textit{Jamie Lannister}, \textit{Tyrion}, \textit{Tormund}, \textit{Missandei}, \textit{Melisandre}, and \textit{Sandor Clagane}, among few others.
The algorithm also discovers a community related to the sub-plotline of \textit{Yara Greyjoy} and some lords loyal to her: after having being saved by her brother \textit{Theon}, she leaves the stage to claim her land. She reappears only in the last episode of the season, together with the other main characters. In this group, we can also find two royal background characters - \textit{Edmure Tully} and \textit{Robin Arryn} - that do not contribute to the development of the main plotline and only appear in the last episode.

\subsubsection{Which are the most important characters?} 
Identifying truly important and influential characters in a vast narrative like GoT may not be a trivial task, as it depends on the considered level of granularity. In these cases, the main character(s) in each plotline is referred with the term \textit{fractal protagonist(s)}, to indicate that the answer to ``who is the protagonist" depends on the specific plotline~\cite{networkofthrones}. 
Following the same methodology of previous experiments, in this section, we focus on evaluating GoT characters according to both the degree and betweenness centrality metrics, exploiting Seasons$\times$Scenes hypergraphs and the corresponding two-section graphs.

\begin{figure}[!h]
%\begin{sidewaysfigure*}[ht]
    \centering
    \begin{subfigure}{.47\columnwidth}
      \centering
      % include first image
      \includegraphics[width=\linewidth]{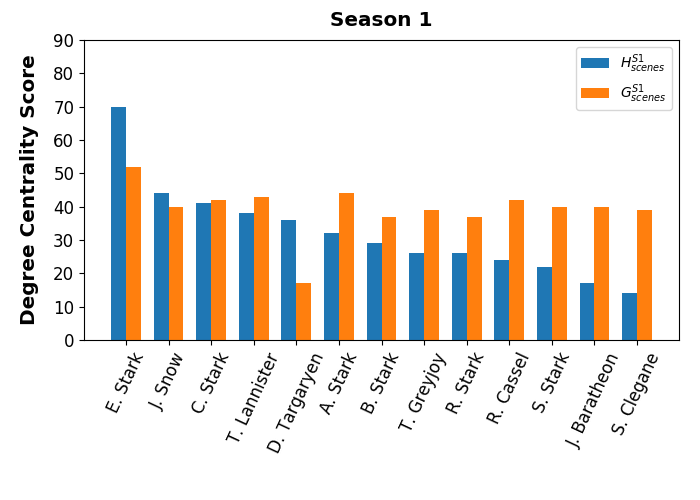} \vspace*{-0.8truecm}  
      %\caption{The hypergraph  $H^{S1}_{episodes}$}
      \label{fig:sub-first}
    \end{subfigure}
    \begin{subfigure}{.47\columnwidth}
      \centering
      % include second image
      \includegraphics[width=\linewidth]{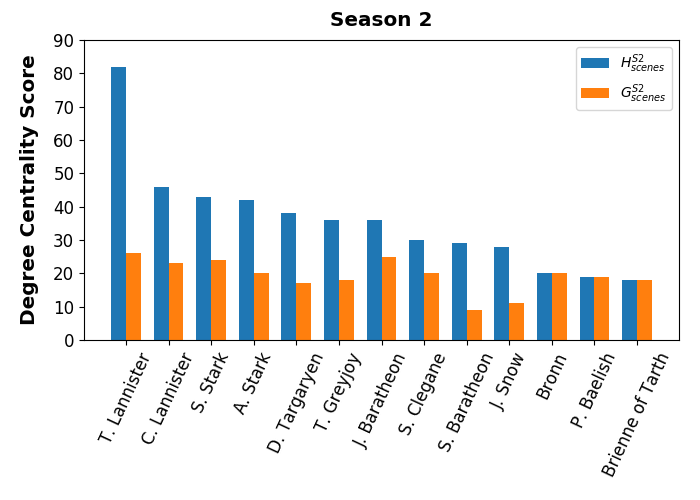} \vspace*{-0.8truecm}
        %\caption{The hypergraph  $H^{S2}_{episodes}$} 
      \label{fig:sub-second}
    \end{subfigure}
    
    \centering
    \begin{subfigure}{.47\columnwidth}
      \centering
      % include first image
      \includegraphics[width=\linewidth]{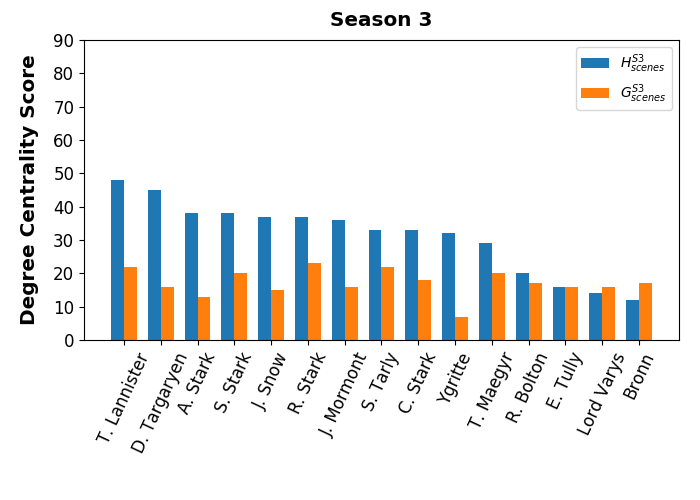}  \vspace*{-0.8truecm}  
      %\caption{The  hypergraph  $H^{S3}_{episodes}$}
      \label{fig:sub-first}
    \end{subfigure}
    \begin{subfigure}{.47\columnwidth}
      \centering
      % include second image
      \includegraphics[width=\linewidth]{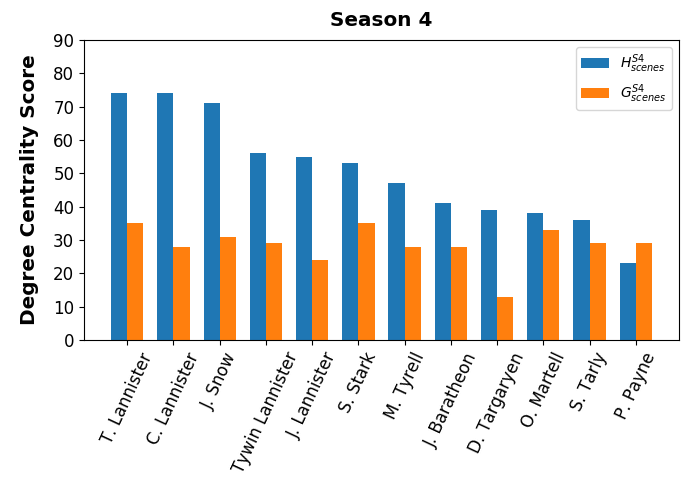}  \vspace*{-0.8truecm}  
       %\caption{The hypergraph  $H^{S4}_{episodes}$}
      \label{fig:sub-second}
    \end{subfigure}

        \centering
    \begin{subfigure}{.47\columnwidth}
      \centering
      % include first image
      \includegraphics[width=\linewidth]{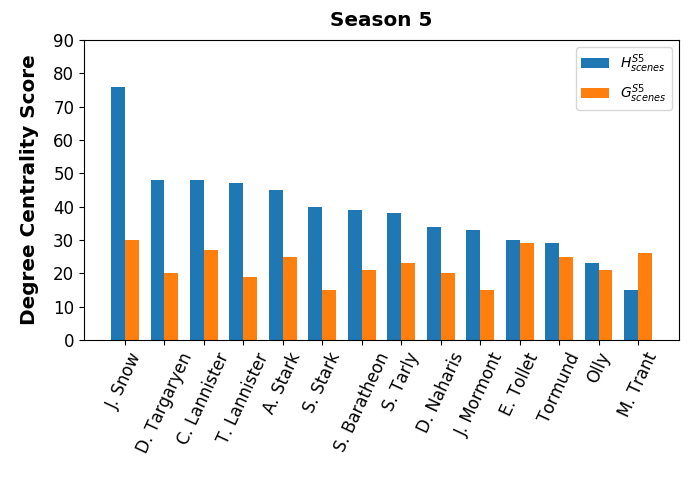}  \vspace*{-0.8truecm}  
      %\caption{The hypergraph  $H^{S5}_{episodes}$}
      \label{fig:sub-first}
    \end{subfigure}
    \begin{subfigure}{.47\columnwidth}
      \centering
      % include second image
      \includegraphics[width=\linewidth]{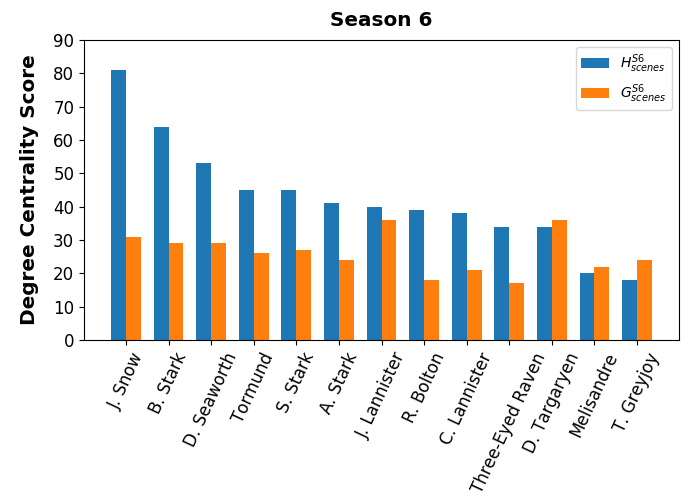}  \vspace*{-0.8truecm}  
  %\caption{The  hypergraph  $H^{S6}_{episodes}$}
      \label{fig:sub-second}
    \end{subfigure}

        \centering
    \begin{subfigure}{.47\columnwidth}
      \centering
      % include first image
      \includegraphics[width=\linewidth]{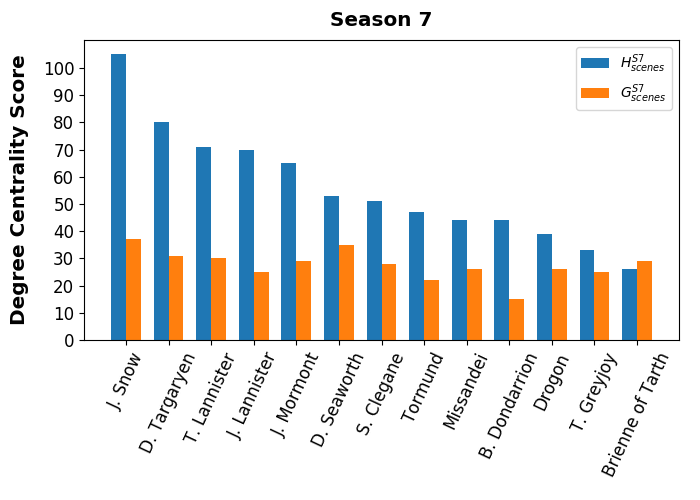}  \vspace*{-0.8truecm}  
     %\caption{The  hypergraph  $H^{S7}_{episodes}$}
      \label{fig:sub-first}
    \end{subfigure}
    \begin{subfigure}{.47\columnwidth}
      \centering
      % include second image
      \includegraphics[width=\linewidth]{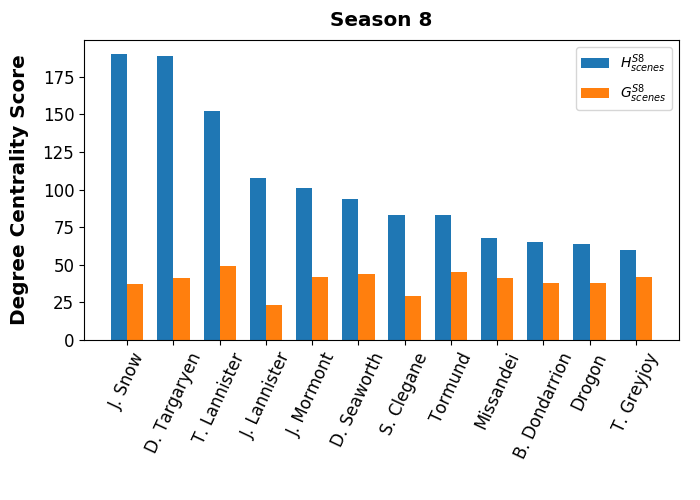}  \vspace*{-0.8truecm}  
     %\caption{The  hypergraph  $H^{S8}_{episodes}$}
      \label{fig:sub-second}
    \end{subfigure}
\caption{GoT characters degree centrality scores per season.} 
\label{fig:got_degree_centrality}
\end{figure}

\paragraph{Degree centrality.}
Generally speaking, this metric gives information about the number of interactions of a node. If we consider a hypergraph $H^s_{scenes}$, the degree centrality of a node is the number of scenes in a given season $s$ a character $v$ appears in. In other words, we are enumerating the number of hyperedges where the vertex is contained. 
Formally,

$$
C^H(v) = |E(v)| 
$$

Analogously, the degree centrality of a character $v$, on the associated two-section graph $G=[H^s_{scenes}]_2$, represents the number of characters he/she played with during season $s$.
Formally,

$$
CG(v) = deg(v) 
$$

Figure \ref{fig:got_degree_centrality} shows that the information provided by the hypergraph analysis is much better to distinguish the centrality of characters. Indeed, each figure depicts the centrality values of the $10$ most important characters of the corresponding season. Only $C^H(v)$ exhibits a clear trend, which is also more coherent among seasons. For instance, \textit{Jon Snow} is the higher degree central character in the last $4$ seasons, while \text{Tyrion Lannister} in the $2$, $3$, and $4$ seasons. 

\paragraph{Betweenness centrality.}
We investigated the importance of the characters also evaluating the betweenness centrality (BC) metric of hypergraph nodes. BC measures the centrality of a node by computing the number of times that a node acts as a bridge between the other two nodes, considering the shortest paths between them. Along the same line of \textit{HyperNetX}\cite{hyperx}, we define the \textit{$s$-node-shortest-path} between two different nodes $u$ and $v$ is the shortest \textit{$s$-node-walk} between them. A $s$-node-walk is a sequence of nodes (characters) such that they share at least $s$ hyperedges. We notice that a $1$-node-walk corresponds to a walk on a graph (or in on the two-section graph), and consequently, a $1$-node-shortest-path is a classical shortest path. Moreover, using the $s$-node-shortest-path definition, we are able to compute the BC considering a path made using more robust connections (or interactions), which we suppose to be more precise to evaluate the importance of the characters in each season.   
Formally, the $s$-betweenness centrality is defined as
$$
C_B^s = \sum_{x \neq v \neq y \in V} \frac{\sigma^s_{xy}(v)}{\sigma^s_{xy}},
$$
where $\sigma^s_{xy}(v)$ is the number of the $s$-node-shortest-paths between two vertices $x$ and $y$ that pass through $v$, while $\sigma^s_{xy}$ is the total number of $s$-node-shortest-paths between $x$ and $y$. We notice that $C_B^s$ generalizes the definition of betweenness centrality. Using $s=1$, $C_B^1$ is the betweenness centrality of the two-section view of $H$.
\begin{figure}[t!]
%\begin{sidewaysfigure*}[ht]
    \centering
    \begin{subfigure}{.47\columnwidth}
      \centering
      % include first image
      \includegraphics[width=\linewidth]{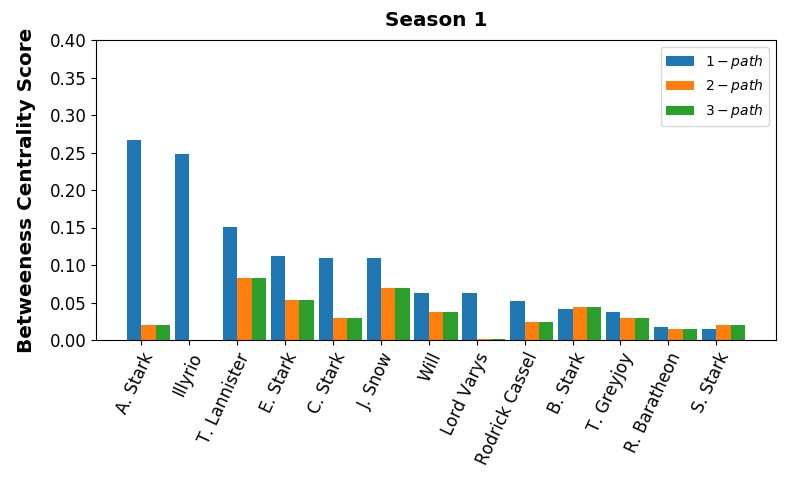} \vspace*{-0.8truecm}  
      %\caption{The hypergraph  $H^{S1}_{episodes}$}
      \label{fig:sub-first}
    \end{subfigure}
    \begin{subfigure}{.47\columnwidth}
      \centering
      % include second image
      \includegraphics[width=\linewidth]{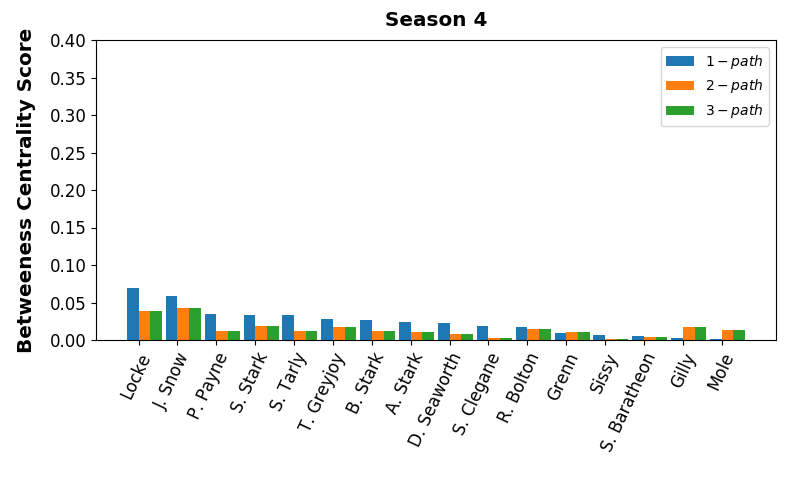} \vspace*{-0.8truecm}
        %\caption{The hypergraph  $H^{S2}_{episodes}$} 
      \label{fig:sub-second}
    \end{subfigure}
\caption{GoT characters $s$-betweenness centrality scores per Season $1$ and $4$.} 
\label{fig:got_betweeness_centrality}
\end{figure}
The rationale behind this generalized definition is that it allows measuring the centrality of the nodes according to a robust shortest path definition. We compared the obtained results varying the value of $s$ from $1$ to $3$,  measuring their correlation to understand whether they provide different information. 
In Table~\ref{tab:pearson-cor}, we report the correlation scores for the season with minimum (season $1$) and maximum (season $4$) correlation. Furthermore, Figure~\ref{fig:got_betweeness_centrality} depicts the corresponding $s$-betweenness centrality values. It is worth mentioning, especially in season $1$, that each character rank varies according to the $s$-value.

\begin{table}[h!]
\centering

\begin{tabular}{lccc}
\toprule
& \multicolumn{1}{c}{$\rho(1, 2)$} & \multicolumn{1}{c}{$\rho(1, 3)$} & \multicolumn{1}{c}{$\rho(2, 3)$} \\
\midrule
\textbf{Season 1 - MIN} & -0.0078 & 0.1174 & 0.8366 \\
\textbf{Season 4 - MAX} & 0.8004  & 0.7962 & 0.7999 \\
\bottomrule
\vspace{5pt}
\end{tabular}
\caption{Pearson correlation values for $s$-betweenness centrality using $\{1,2,3\}$-path on the seasons with minimum and maximum correlation. \label{tab:pearson-cor}}
\end{table}

\section{Conclusions}\label{sec:conclusion}
In this work, we have presented a new library for analyzing, exploring, and visualizing of hypergraph structures through an optimized set of hypergraph manipulations. The library, named \texttt{SimpleHypergraphs.jl}, provides Hypergraph views built exploiting the popular Julia library \texttt{LightGraphs.jl} for manipulating graphs. Functionalities for the I/O, manipulation, transformation, and visualization of hypergraphs have been developed and are available on a public GitHub repository.
The library enables the user defining meta-information type as well as attaching meta-data values of arbitrary nature to both vertices and hyperedges. This approach allows programmers to efficiently analyze the structural properties of the network, combined with the possibility to perform semantic analysis based on the attached meta-data. To the best of our knowledge, \texttt{SimpleHypergraphs.jl} is the only hypergraph software tool providing this functionality.

Hypergraphs are a natural generalization of graphs and, at least theoretically, they provide a much richer structure than their well-known graph counterparts. More importantly, they seem to be more suitable than graphs to model many natural phenomena that involve group-based interactions, such as collaborative activities. However, it is not still clear if the advantage of preserving a more detailed relationship structure justifies a more complicated data structure and, as a consequence, more complex underlying algorithms. 
In our work, we investigated this aspect of the research on hypergraphs by providing two case studies belonging to different domains. In the first case study, we explored the application of data analysis using hypergraphs for understanding users reviewing activities on the social network\textit{Yelp.com}. In the second case study, we proposed the application of hypergraph analysis for discovering the community structure of a network society based on the interactions of characters in the \textit{Game of Thrones} TV series. Results suggest that the hypergraph structure seems to improve the analysis of networks defined by \textit{many-to-many} relationships, as they convey more information compared to the alternative view using the graph structure.

Future investigations are necessary to validate the conclusions drawn from this study. We plan to explore other types of data by widening the application domain. Furthermore, we are currently working on improving the library functionalities by implementing more algorithms and different accurate and engaging visualizations mechanisms.

\clearpage
%\nocite{*}
\bibliographystyle{acm}

\end{document}